\documentclass[%
reprint,
superscriptaddress,
notitlepage,
amsmath,amssymb,
aps,
onecolumn,
floatfix,
tightenlines,
longbibliography,
11pt,
]{revtex4-1}

\usepackage{psfrag,graphicx,epsfig,color}% Include figure files
\usepackage{dcolumn}% Align table columns on decimal point
\usepackage{bm}% bold math
\usepackage{natbib}
\usepackage{float}
\usepackage[usenames,dvipsnames,svgnames,table]{xcolor}
\usepackage{subfigure}
\usepackage[normalem]{ulem}

\def\epsav {{ \langle \epsilon \rangle }}
\def\ktime {{\tau_\eta}}

\def\uu {{\mathbf{u}}}

\def\ff {{\mathbf{\Phi}}}
\def\bi {\begin{itemize}}
\def\ei {\end{itemize}}

\begin{document}

\title{Single-particle Lagrangian statistics from 
direct numerical simulations \\ of rotating-stratified turbulence}

\author{D. Buaria}
\email[]{dhawal.buaria@ds.mpg.de}
\thanks{(present address: New York University, NY 11201, USA; dhawal.buaria@nyu.edu)}
\affiliation{Max Planck Institute for Dynamics and Self-Organization, 37077 G\"ottingen, Germany}
\author{A. Pumir}
\affiliation{Laboratoire de Physique, ENS de Lyon, CNRS and Universit\'e de Lyon, 69007 Lyon, France}
\affiliation{Max Planck Institute for Dynamics and Self-Organization, 37077 G\"ottingen, Germany}
\author{F. Feraco}
\affiliation{Laboratoire de M\'ecanique des Fluides et d'Acoustique, \'Ecole Centrale de Lyon, CNRS, \\ 
Universit\'e Claude Bernard Lyon 1, INSA de Lyon, 69134 \'Ecully, France}
\affiliation{Dipartimento di Fisica, Universit\'a della Calabria, 87036, Arcavacata di Rende, Italy}
\author{R. Marino}
\affiliation{Laboratoire de M\'ecanique des Fluides et d'Acoustique, \'Ecole Centrale de Lyon, CNRS, \\ 
Universit\'e Claude Bernard Lyon 1, INSA de Lyon, 69134 \'Ecully, France}
%\affiliation{Universit\'e de Lyon, \'Ecole Centrale de Lyon, Universit\'e Claude Bernard, CNRS, INSA de Lyon, 
%Laboratoire de M\'ecanique des Fluides et d'Acoustique, 69135, \'Ecully, France}
\author{A. Pouquet}
\affiliation{Laboratory for Atmospheric and Space Physics, University of Colorado \\
and National Center for Atmospheric Research, Boulder, CO 80309, USA}
%\affiliation{Laboratory for Atmospheric and Space Physics, University of Colorado, Boulder, CO 80309, USA}
%\affiliation{National Center for Atmospheric Research, Boulder, CO 80307, USA}
\author{D. Rosenberg}
\affiliation{288 Harper St., Louisville, CO 80027, USA}
\author{L. Primavera}
\affiliation{Dipartimento di Fisica, Universit\'a della Calabria, 87036, Arcavacata di Rende, Italy}

\date{\today}

\begin{abstract}

Geophysical fluid flows are predominantly turbulent and often 
strongly affected by the Earth's rotation, 
as well as by stable density stratification. 
%of acceleration and particle transport,
Using direct numerical simulations of forced Boussinesq equations,
we study the influence of these effects on the motion of fluid particles.
We perform a detailed study of Lagrangian statistics of acceleration, velocity
and related quantities,
focusing on cases where 
the frequencies associated with rotation and stratification (RaS),
$f$ and $N$ respectively,
are held at a fixed ratio $N/f=5$. 
The simulations are  performed in a periodic domain,
at Reynolds number $Re \approx 4000$,
and Froude number $Fr$ in the range
$0.03 \lesssim Fr \lesssim 0.2$
(with Rossby number $Ro=5Fr$).
%\{Due to \{the known} anisotropy 
%\{induced by rotation and stratification}, we consider separately
%the motion in the horizontal and vertical directions.}
%Both stratification and the axis of rotation are aligned
%in the vertical direction, and their associated 
%with stratification being comparatively stronger than rotation.
%The simulations are  performed in a periodic domain,
%at Reynolds number $Re \approx 4000$,
%and Froude number $Fr$ in the range
%$0.03 \lesssim Fr \lesssim 0.2$
%(and Rossy number $Ro$ being $5Fr$).
As the intensity of RaS increases, 
a sharp transition is observed between a regime dominated 
by eddies to a regime dominated by waves,
which corresponds to $Fr \lesssim 0.07$.
For the given runs, this transition to a wave-dominated regime 
can also be seemingly described by
simply comparing the time scales $1/N$  and $\tau_\eta$,
the latter being the Kolmogorov time scale 
based on the mean kinetic energy dissipation.
Due to the known anisotropy 
induced by RaS, we consider separately
the motion in the horizontal and vertical directions.
In the regime $N\tau_\eta < 1$,
acceleration statistics 
exhibit well known characteristics of isotropic
turbulence in both directions,
such as probability density functions (PDFs) with wide tails
and acceleration variance approximately scaling as per Kolmogorov's theory.
In contrast for $N\tau_\eta > 1$, they 
behave very differently, experiencing the direct influence
of the imposed rotation and stratification.
On the other hand, the Lagrangian velocity statistics
exhibit visible anisotropy for all runs; nevertheless
the degree of anisotropy becomes very strong in the regime
$N\tau_\eta > 1$.
%In considering the mean square displacements of particles, we find
We observe
that in the regime $N\tau_\eta < 1$, rotation enhances the 
mean-square displacements in horizontal
planes in the ballistic regime at short times, but
suppresses them in the diffusive regime at longer times.
This suppression of the horizontal displacements becomes 
stronger in the regime $N\tau_\eta>1$,
with no clear diffusive behavior.
In contrast, the displacements in the vertical direction are always
reduced. This inhibition is extremely strong 
in the $N\tau_\eta > 1$ regime, leading
to a scenario where particles almost appear to be trapped in 
horizontal planes. 
\end{abstract}

\maketitle

\section{Introduction}

Transport of material substances plays a crucial role in  
many geophysical processes \cite{pedlosky,weiss07}, e.g. 
dispersion of pollutants
and contaminants \cite{wyngaard,  stohl02},
droplet dynamics in clouds \cite{shaw_03}, 
mixing of planktons and other biomatter in the oceans \cite{guasto}.
The Lagrangian viewpoint
following the motion of fluid particles \cite{MY.I}
or analogous entities such as Brownian or inertial particles
\cite{BYS.2016, pumir16},
provides a natural description of such transport processes.
%or analogous entities such as Brownian particles or inertial particles
%\cite{BYS.2016, pumir16}.
Not only are most geophysical flows turbulent,
they are also often strongly influenced
by anisotropies due to effects such as rotation 
(Coriolis force) and stratification (buoyancy force) 
or presence of magnetic fields \cite{Davidson_13}.
While particle dispersion has been studied extensively
in isotropic turbulence, e.g. see reviews 
\cite{Sawford:01, Yeung02, TB_ARFM}, 
it has only recently started receiving
attention in anisotropic flows. In particular, 
several studies have focused on flows considering either the effects
of rotation 
\cite{DelCastello+:11a,DelCastello+:11b, Naso:2012}
or stratification separately 
\cite{Artrijk:08, Brethouwer:09, Sujovolsky:18}. 
However, in many applications, adequately describing the
observed flow physics necessitates examining the combined effects of
rotation and stratification (RaS), 
%Whereas either rotation or stratification may prevail in some applications,
%it is often necessary to take into account 
%their combined effects,
e.g. in the southern abyssal 
oceans with particularly high mixing intensities
\cite{Naveira:04,Nikurashin:12,vanharen_16j}.
%In instance, the global energy budget
%is often derived by assuming a
%geostrophic balance between 
%pressure gradient, buoyancy and Coriolis forces
%at the large scales (cite?).

%We are concerned here with the effect of the small scales of the turbulent 
%motion on mixing. 
Arguably the most challenging aspect of studying the
combined effects of RaS in turbulence is the 
enormous range 
of spatial and temporal scales associated with such flows.
In isotropic turbulence, typically the only 
governing parameter is the
Reynolds number ($Re$), which directly provides
a measure of the range of scales in the flow. 
However, at least three additional parameters
have to be considered in RaS turbulence,
namely, the Rossby number ($Ro$) and the Froude
number ($Fr$) which respectively measure the strength of 
RaS, and the
Prandtl number ($Pr$) which is the 
ratio of the fluid viscosity to the thermal diffusivity
(see Section~\ref{sec:numerics} for precise definitions).
This leads to consideration of additional
length and time scales 
-- beyond the already wide range existing
in isotropic turbulence --
and makes the flow dynamics far more involved. In particular,
it has been recognized that the interaction between linear 
and non-linear processes in such flows leads to 
a rich variety of complex behavior, 
such as spontaneous generation of helicity, 
dual cascading of energy and many more \cite{marino13b,Marino:13,haller_arfm}. 
Advances in computing power in the last decade, has allowed 
for significant progress in understanding such flows, although
mostly from the Eulerian perspective 
\cite{ivey_08,deusebio_14b,Rosenberg:15,lindborg_17,waite_17,gregg_18}.

In this work,
utilizing direct numerical simulations (DNS) of Boussinesq equations,
our objective is to investigate the dispersion of
fluid particles in RaS turbulence.
In particular, we focus on the motion of single individual particles.
In view of the very large parameter space associated with 
RaS turbulence, we limit our investigation
to the case where $Ro/Fr=5$, i.e., the strength of stratification is
five times that of rotation, which is largely relevant to the southern abyssal 
oceans~\cite{Naveira:04,Nikurashin:12}.  
We use the same grid sizes for all cases,
giving $Re \approx 4000-5000$, and 
Froude numbers in the range $0.03 \lesssim Fr \lesssim 0.2$.
This regime of $Fr$ 
corresponds to the transitional regime, where both turbulent eddies
and inertia-gravity waves are 
expected to play an important role \cite{pouquet_18}. 
As recently demonstrated in the case of purely stratified flows, 
the transition towards a wave
dominated regime is accompanied by 
intermittent large-scale vertical drafts \cite{Rorai:14,Feraco:18}. 
We show that the presence of additional rotation
does not, in general, prevent the
manifestation of this intermittency and thereby also
investigate its effect on Lagrangian statistics.

In studying
the motion of single particles,
in the spirit of earlier studies
\cite{Sawford:01, Yeung02, TB_ARFM},
we investigate
the properties of both their acceleration and
velocity.
The intrinsic anisotropy of the flow makes it necessary
to distinguish throughout between motion in the horizontal plane and in
the vertical direction, which are respectively perpendicular and parallel
to the direction of imposed stratification (and also the axis of rotation).
However, the anisotropy of acceleration and velocity reveal
%{, which are induced by different aspects of the turbulent motion,} 
strikingly different properties.
For acceleration statistics, which are reflective of small-scales
of turbulence, two distinct regimes are observed depending on the 
strength of imposed RaS.
At weak or moderate RaS, the properties of acceleration
are qualitatively similar to those documented in isotropic turbulence
(with minor quantitative deviations), suggesting that
the imposed anisotropy at large scales does not significantly affect the small-scales.
%We find two main types of behavior, depending on the relative
%importance of stratification and rotation, on the one hand, and of
%the turbulent eddies, on the other hand. 
%At moderate values of stratification and rotation, the properties of
%the velocity and acceleration of tracer particles are qualitatively 
%similar to those documented in isotropic turbulence, 
%with minor deviations.
On the other hand, when RaS becomes very strong, 
acceleration statistics are significantly affected
and exhibit striking differences between 
the horizontal and vertical motions. 
In the range of parameters considered here, we observe a sharp transition
between the two regimes, which can be simply quantified with
the ratio
of the Kolmogorov time scale and the stratification time scale
(which is simply the inverse of the Brunt-V\"ais\"ala frequency).
The qualitative change in the statistical properties of acceleration 
occurs simultaneously with the appearance of intermittent bursts in the 
flow.

The Lagrangian velocity statistics, in contrast to acceleration,
are always affected by the imposed RaS, with the degree
of anisotropy expectedly increasing with the strength of the imposed RaS.
However, the transition observed for acceleration statistics 
is also visible in velocity statistics, resulting
in even stronger anisotropy in velocity statistics
when RaS is very strong. 
Accordingly, we characterize the integral time scales based on velocity 
autocorrelations and investigate the displacement of particles.
The particles move ballistically at short times.
At long times, the emergence of a diffusive regime
is evident for runs with moderate RaS. 
However for strong RaS, the dispersion, particularly
in vertical direction is strongly suppressed,
qualitatively consistent with earlier works on 
purely stratified flows.

This paper is organized as follows.
In Section \ref{sec:numerics}, we discuss our numerical methods.
Essential features of the underlying Eulerian flow are discussed in
Section~\ref{sec:euler}.
The statistics of acceleration are presented in Section \ref{sec:acceleration}.
Section \ref{sec:velocity} presents our results on the Lagrangian
autocorrelation functions of velocity along
with results on how particles are displaced by the flow.
Finally, Section \ref{sec:concl} contains our concluding remarks.

\section{Numerical Method and Database}
\label{sec:numerics}

We simulate the Eulerian flow by numerically integrating the 
incompressible Boussinesq equations 
in a rotating frame,
with constant solid body rotation rate $\Omega$ (and frequency $f = 2 \Omega$) 
and gravity $g$ anti-aligned in the vertical ($z$) direction:
\begin{align}
\partial \uu /\partial t  + \uu \cdot \nabla \uu &=
-\nabla P - f \mathbf{e}_z \times \uu - N \theta \mathbf{e}_z + \nu \nabla^2 \uu + \ff   \label{eq:B1} \\ 
\partial \theta /\partial t  + \uu \cdot \nabla \theta &=
N w + \kappa \nabla^2 \uu                                                                \label{eq:B2} \\
\nabla \cdot \uu &= 0                                                                         \label{eq:B3}
\end{align}
Here, $\uu=(u,v,w)$ is the velocity field, $\theta$ is the
temperature fluctuation (in units of velocity), $P$ is the pressure normalized
by the background density,
$\nu$ is the kinematic viscosity, $\kappa$ is the thermal diffusivity
and $N$ is the  Brunt-V\"ais\"ala frequency, which characterizes the strength
of imposed stratification. 
We take $\nu = \kappa$, assuming the Prandtl number to be unity.
A large-scale stochastic forcing term, $\ff$, is utilized to achieve and maintain a 
statistically stationary state. 
The forcing is random in time and isotropic in Fourier space,
with the energy being injected in a spherical shell of
wavenumbers given by $2 < |\mathbf{k}| < 3$ \cite{Marino:13a},
with the characteristic forcing length scale 
$L_f = 2 \pi/ 2.5$.
While the use of isotropic forcing to simulate
innately anisotropic flows may appear unphysical in the present
context, earlier comparisons
with a quasi-geostrophic forcing
have demonstrated that the precise nature of
the forcing does not have any significant effect on the flow properties
in the stationary state, 
at least in the regimes studied here \cite{Marino:13,Rosenberg:15,pouquet_18,pouquet_19e}.

%While the use of large-scale isotropic forcing to simulate 
%innately anisotropic flows may appear unphysical,
%our results shown in following sections demonstrate that
%are notoriousl anisotropic, may appear as artificial. The isotropy of the
%forcing used here does not prevent the development of very anisotropic flow 
%properties, as documented in the Sections~\ref{sec:euler} 
%(see Table~\ref{table:stat_v})
%and \ref{sec:acceleration}, or the formation of vertical
%structures at scales smaller than $L_f$. 

To characterize RaS, we introduce 
the dimensionless Rossby ($Ro$) and Froude ($Fr$) 
numbers defined as~\cite{Davidson_13}:
\begin{equation}
Ro = \frac{U}{L f} \ , \ \   Fr = \frac{U}{L N}
\label{eq:def_Ro_Fr}
\end{equation}
where $L$ and $U$ are respectively the characteristic length 
and velocity scales of the large-eddies of the flow,
defined as $L = L_f$,  and 
$U = \langle |\mathbf{u}|^2 \rangle^{1/2}$,
the mean
amplitude of the velocity fluctuations. 
An important parameter is the ratio $N/f$ ($=Ro/Fr$), 
which measures the relative strength 
of RaS.
As already mentioned, we maintain $N/f=5$ for 
all runs, 
as relevant in some oceanographic
situations~\cite{Naveira:04,Nikurashin:12}.

In addition, we define the following dimensionless numbers
\begin{equation} 
Re = \frac{U L}{\nu} \ , \ \  R_{IB} = \frac{\langle \epsilon \rangle}{\nu N^2}  \ , \ \ R_{\cal B} = 
Re Fr^2   \, \ , \ Ro_{\omega} = 
\frac{\langle\omega^2\rangle^{1/2}}{f} \ ,
\label{eq:def_Reyn}
\end{equation}
where $\langle \epsilon \rangle$ is the 
mean dissipation rate of turbulent kinetic energy
and $\langle \omega^2 \rangle$ is the mean enstrophy
density (with $\omega$ being the magnitude of vorticity). 
It is worth noting that 
in homogeneous turbulence, as considered in this work,
the two are related as
$\langle\omega^2\rangle = \langle \epsilon\rangle/\nu = 1/\tau_\eta^2$,
where 
\begin{align}
\tau_\eta=(\nu/\langle \epsilon \rangle)^{1/2}
\end{align}
is the Kolmogorov time scale
characterizing the motion of turbulent eddies \cite{BPBY2019}.
Here, $Re$ is the Reynolds number based on large scale.
The parameter $R_{IB}$, often referred to as the
buoyancy Reynolds number, provides a  
measure of the relative importance of the waves induced
by stratification with respect to 
turbulent eddies in the flow (or alternatively,
the relative importance of waves and non-linear processes), e.g. see
\cite{marino_15w,ivey_08}.
This is evident by considering 
the Ozmidov length scale 
$\ell_{OZ} = (\langle \epsilon \rangle/N^3)^{1/2}$,
which characterizes the gravity waves
and the Kolmogorov length scale, $\eta = (\nu^3/\langle \epsilon \rangle)^{1/4}$,
which characterizes the dissipation scale of turbulence,
thereby giving $R_{IB} = (\ell_{OZ}/\eta)^{4/3}$. 
Alternatively, $R_{IB}$ can be written as the ratio
of two time scales: $R_{IB} = (T_N/\tau_\eta)^2$, where
$T_N=1/N$ is the stratification time scale
(and  $\tau_\eta$ is
the Kolmogorov time scale), once again characterizing the
relative importance of stratification to that of turbulent eddies.
An alternative perspective is also offered by the asymptotic scaling analysis 
of \cite{Brethouwer:07}, which suggests that $R_{IB}$ 
can be interpreted as a replacement for the classical
Reynolds number in strongly stratified turbulence -- providing
a relative measure of inertial and viscous forces.
However, utilizing this interpretation in presence of
rotation requires caution, since it invalidates the said asymptotic analysis
(especially when rotation and stratification are comparable in strength).

A separate definition of the buoyancy Reynolds 
number~\cite{Rosenberg:15,pouquet_18}
is given by the parameter $R_{\cal B}$. 
The two parameters are related as $R_{IB} = \beta R_{\cal B}$, where 
$\beta = \langle \epsilon \rangle L/U^3$. In fully developed turbulence, 
$\beta$ is expected to be 
constant as a consequence of the dissipation anomaly \cite{Ishihara09}.
However, in the regimes considered here, it can be shown that $\beta$  
depends on the strength of stratification \cite{pouquet_18}.
Hence in our analysis, we utilize $R_{IB}$ 
to quantify the relative strength of stratification and turbulence.
The relevance of this choice 
will become evident as our results are presented in later sections.
Finally, in addition to the Rossby number $Ro$ based on large scales, 
it is useful to define the micro-Rossby number $Ro_{\omega}$,
which in contrast to $R_{IB}$, measures the relative strength of 
small-scale turbulent motions to that of the imposed rotation.
However, using $\langle \epsilon \rangle = \nu \langle \omega^2\rangle$ due to
statistical homogeneity,
$Ro_{\omega}$ can be simply related to 
$R_{IB}$ as:
$Ro^2_{\omega}/R_{IB} = (N/f)^2$.
Since $N/f$ is held constant in this work,
the parameters
$Ro_{\omega}$ and 
$R_{IB}$ essentially provide the same information.

The database utilized in the current work,
along with the main simulation parameters, is
summarized in Table~\ref{tab:param}.
The simulations were carried out using the
Geophysical High-Order Suite for Turbulence (GHOST) code, a versatile,
highly parallelized, pseudo-spectral code, 
utilizing hybrid MPI-OpenMP programming model, 
with second order
explicit Runge-Kutta time stepping \cite{Mininni:11}. 
All runs correspond to a $(2\pi)^3$ periodic domain with $512^3$ grid points.
The parameters 
$N$ and $f$ are varied over a range of values 
keeping the ratio $N/f$ fixed at 5. 
In addition, for a systematic comparison, 
we have also performed  additional simulations
by setting $N=0$ or $f=0$ or both. 
The simulation with $N=f=0$ (case 0)
simply corresponds to homogeneous isotropic turbulence (HIT).
All the runs reported were started from an initial condition, consisting of
a few random modes in the velocity field, 
whereas the temperature field, $\theta$, was initialized to zero.
The Boussinesq equations 
were integrated until a statistically stationary state was reached.
For adequate statistical sampling, we considered a simulation period of
at least $10-20T_E$ in the stationary
state for each case, where 
$T_E=L/U$ is the large eddy turnover time.
(Some additional details are available in the Supplementary 
Material \cite{supp}).

\begin{table}[h]
 \centering 
 \begin{tabular}{l|c|cccccc|cc}
  \hline \hline
    case & 0 (HIT)      & 1      & 2      & 3     & 4     & 5     & 6     & 7 (s)     & 8 (r)     \\
\hline
    $N$    & 0        & 2.948  & 4.915  & 7.372 & 11.80 & 14.74 & 16.62 & 14.74    & 0         \\
    $f$    & 0        & 0.5896 & 0.9830 & 1.474 & 2.360 & 2.948 & 3.320 & 0        & 2.948     \\
    $\nu$  & 0.0015   & 0.001  & 0.001  & 0.001 & 0.001 & 0.001 & 0.001 & 0.001    & 0.001     \\
\hline
    $Fr$   & $\infty$ & 0.168  & 0.114  & 0.086 & 0.069 & 0.063 & 0.045 & 0.030    & $\infty$  \\
    $Ro$   & $\infty$ & 0.840  & 0.570  & 0.430 & 0.345 & 0.315 & 0.225 & $\infty$ & 0.140     \\
    $Re$   & 2379     & 3117   & 3537   & 3963  & 5199  & 5861  & 4744  & 3942     & 2645      \\
    $R_{\cal B}$  & $\infty$ & 90.1   & 42.9   & 26.9  & 24.7  & 23.2  & 9.61  & 3.0      & $\infty$  \\
\hline
$\langle \epsilon \rangle$  & 0.375      & 0.198    & 0.182    & 0.146   & 0.059  & 0.049  & 0.027  & 0.050     & 0.065   \\
$R_{IB}$   & $\infty$      & 22.8    & 7.53    & 2.69   & 0.422  & 0.226  & 0.096  & 0.229     & $\infty$   \\
$Ro_\omega$& $\infty$ & 23.9 & 13.7 & 8.22 & 3.25 & 2.38 & 1.55 & $\infty$ & 2.73   \\
\hline   
\end{tabular} 
\caption{
DNS database used in the current work, with corresponding simulation parameters.
All rotating-stratified runs (cases 1-6) correspond  to $N/f=5$. 
The cases 7 and 8, respectively with only stratification and only rotation,
correspond to the $N$ and $f$ values of case 5.
The case 0 corresponds to homogeneous isotropic turbulence (HIT)
with $N=f=0$. 
The mean dissipation of kinetic energy $\langle \epsilon \rangle$
was obtained by averaging over all the grid points and 
the time over which particles were tracked in the simulations.
Other parameters are defined in Eqs.~\eqref{eq:def_Ro_Fr}
and \eqref{eq:def_Reyn}. 
} 
\label{tab:param}
\end{table}

In Table~\ref{tab:param}, 
an important point to note is
that the values of $R_{\cal B}$ and $R_{IB}$ 
monotonically decrease as 
$N$ and $f$
are increased. Values of $N$ and $f$ higher than those shown are
avoided, since the resulting regime would be completely dominated by waves
\cite{pouquet_18}.
The simulations with either $N=0$ or $f=0$ have their  
$f$ or $N$ value (respectively) corresponding to case 5, with one of
the largest $N$ and $f$ values.
However, as evident from Table~\ref{tab:param} the values of the $Fr$ (and $Ro$)
numbers corresponding to runs 
5 and 7 (respectively runs 5 and 8) differ significantly, despite identical
values of the stratification (respectively rotation rate). 
These variations
are a consequence of our dynamical definition of $U$, which 
accounts for
the subtle interplay between turbulence and RaS.
In particular, they reflect the 
significant variation in $U$ itself, which is 
discussed in Section~\ref{sec:euler}. 
Importantly,
we observe that the value of $R_{IB} $ is lower than $0.5$ for Runs 4-6,
as well as for the purely stratified run, 7, suggesting a dominant role
of the waves in these runs. In contrast, $R_{IB} > 2$ in 
all other runs, possibly pointing to a prevalent role of the eddies. 
This leads to two qualitatively different behaviors, as demonstrated by 
our results in the following sections.

Finally, in order to obtain Lagrangian statistics, 
the fluid particles are tracked in time along with the Eulerian flow
in the stationary state,
according to the basic equation of motion
\begin{equation}
\frac{d \mathbf{x^+}(t)}{dt}  = \mathbf{u^+}(t) = \mathbf{u}( \mathbf{x^+}(t), t)
\label{eq:dxdt}
\end{equation}
where the superscript $+$ denotes a Lagrangian quantity and the fluid particle
velocity is simply defined to be the Eulerian velocity field
evaluated at the instantaneous particle position.
For each run listed in Table~\ref{tab:param}, we additionally tracked
the motion of 1.5M particles (M=$10^6$), which are 
randomly distributed
throughout the entire domain at the time of injection. 
The number of fluid particles
are held constant for all cases, since given the same grid size
and approximately similar Reynolds numbers, their sampling requirements
are also approximately the same. 
Similar to the Eulerian grid, the particles are distributed among
parallel processors and tracked in time together with the velocity field
using a second order
Runge-Kutta scheme, and standard interpolation methods based
on cubic splines~\cite{hinsberg, buaria.cpc}.

\section{Eulerian velocity field}
\label{sec:euler}

In this section, we provide a brief overview of the underlying
Eulerian flow, which establishes
the necessary framework for better understanding the Lagrangian statistics
reported subsequently.

\paragraph*{Energy spectra:}

\begin{figure}[h]
\begin{center}
\includegraphics[width=7.9cm]{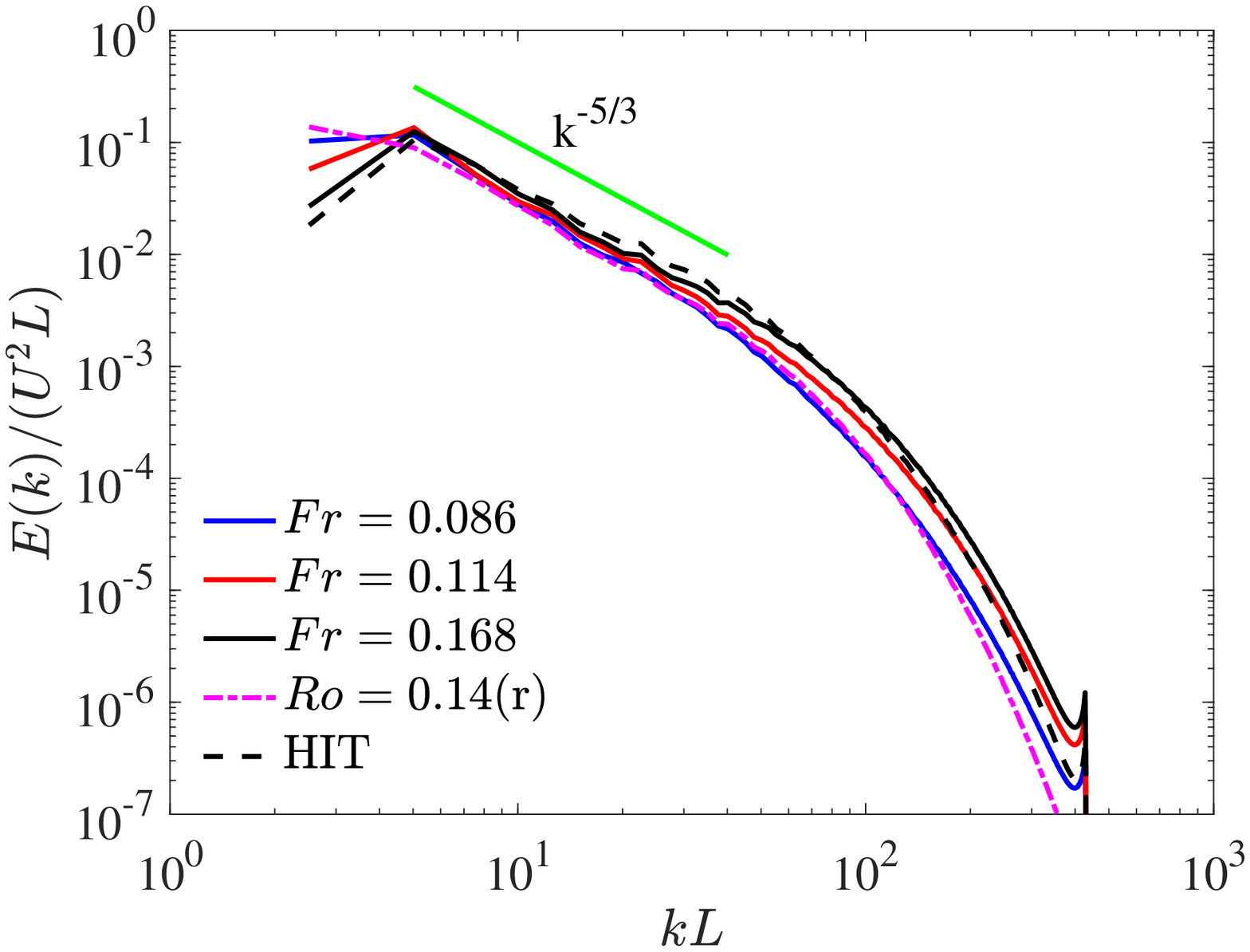} \ \ \ \
\includegraphics[width=7.9cm]{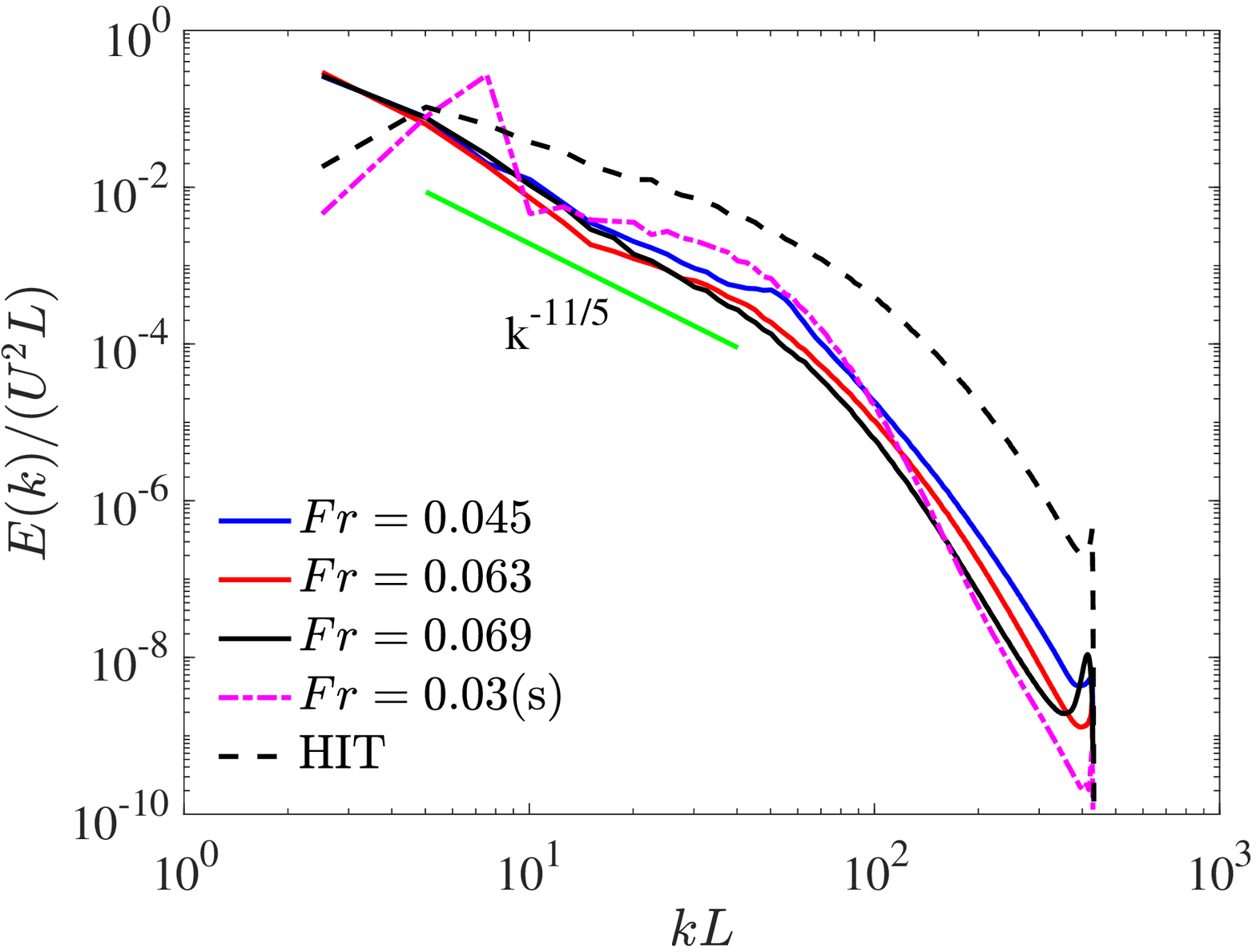}
\caption{
The kinetic energy spectrum $E(k)$ normalized by $U^2L$
as a function of $kL$, where $U$ and $L$ are the large scale
velocity and length scales respectively.
The curves correspond to cases listed in Table~\ref{tab:param}.
%Panels (a) and (b) correspond to the same set of runs
%as in Fig.~\ref{fig:Z_vs_t}(a) and (b) respectively.
%{Panel (a) corresponds to the runs with low RaS, runs 1-3 
%($Fr \ge 0.08$), along with the HIT run (0) and the purely rotating flow (run 8). Panel (b) corresponds to larger RaS, runs 4-6 ($Fr \le \s{0.07}$), along with
%the purely stratified flow (run 7).} 
Solid green lines indicating a spectral slope of 
$k^{-5/3}$ and $k^{-11/5}$ in (a) and (b) respectively
are drawn for reference.
}
\label{fig:spctr}
\end{center}
\begin{picture}(0,0)(0,0)
\thicklines
\put(-55,225){$(a)$}
\put(180,225){$(b)$}
\end{picture}
\end{figure}

We begin by considering the kinetic energy spectra,
as shown in Fig.~\ref{fig:spctr},
which is one of the most common descriptor 
of Eulerian dynamics in turbulent flows \cite{Davidson_13}.
The curves are normalized using the large scale variables $U$ and $L$,
and split into two sub-figures depending on the strength of 
weak rotation and stratification (RaS). 
%We begin by considering the kinetic energy spectra, 
%{as considered \textit{e.g.} in
%\cite{Davidson_13,Brethouwer:07,Rosenberg:15,waite_17,pouquet_18}.
%Fig.~\ref{fig:spctr} shows the kinetic energy spectra,}
%normalized using the large scale variables $U$ and $L$,
%and split into two sub-figures depending on the strength of 
%stratification.
Fig.~\ref{fig:spctr}a shows the spectra for the runs with relatively 
weak RaS,
corresponding to $Fr \ge 0.086$ and  $R_{IB}>1$, 
including the HIT and rotation-only runs.
The main observation is that all the spectra
demonstrate an approximate $k^{-5/3}$ Kolmogorov-like scaling
in an intermediate range of $k$. 
This expectation is justified for the HIT run, 
as well as the $R_{IB} > 1$ runs, for which there is a 
sufficient scale separation between $\ell_{OZ}$ and $\eta$
(note $R_{IB} = (\ell_{OZ}/\eta)^{4/3}$ as discussed earlier),
to recover Kolmogorov-like
turbulence \cite{lindborg_06,Brethouwer:07};
although systematic deviations from the scaling 
are clearly evident as $R_{IB}$ decreases. 
In fact, as shown later in this section,
the large scales for $R_{IB} > 1$ runs are still 
predominantly anisotropic, though their impact
on small-scales is weak.

On the other hand, in Fig.~\ref{fig:spctr}b,
the spectra for runs with stronger stratification,
corresponding to $Fr \le 0.069$ and  $R_{IB}<1$
are shown. The spectrum for the HIT run (dashed black line) 
is also included for comparison.
As evident, this regime significantly differs
from that at weaker stratification,
with the spectra exhibiting a spectral slope steeper
than $k^{-5/3}$
(since $R_{IB} <1$ and hence $\ell_{OZ} < \eta$ precludes
Kolmogorov's theory from being applicable).
It has been observed
in the closely related context of a time-dependent 
flow~\cite{Rosenberg:15,verma_17} that
the spectra followed the Bolgiano-Obhukov prediction \cite{Bolgiano:59, obukhov_59}.
In this regime, energy transfer
is strongly affected by the temperature fluctuations (potential energy).
The straight line in Fig.~\ref{fig:spctr}b
indicates the corresponding $k^{-11/5}$ prediction.
Given the small range of spatial scales available in
our simulations, however, 
it is difficult to conclusively demonstrate that the spectra shown  
indeed conform to a $k^{-11/5}$ scaling
(and the fact that the precise exponent can depend on the Froude number \cite{kimura12}).
Nevertheless, the change in spectral slopes clearly evidences
that the energy transfer between
scales is strongly perturbed in runs with $R_{IB}<1$, and much less so 
in runs with $R_{IB}>1$.
In fact, this demarcation based on $R_{IB}$ becomes increasingly prominent
as more results are discussed in later sections. 
The spectra of the temperature fluctuations (not shown)
follow similar trends as that of velocity.

\paragraph*{Anisotropy of velocity fluctuations:}

The above transition between strong hydrodynamic turbulence and regimes
dominated by RaS, as evidenced from
the energy spectra, clearly warrant further inspection.  
An important aspect of the flows in 
this study is the existence of a strong anisotropy of velocity 
fluctuations. Namely, the presence of RaS induces
very different properties of the velocity fluctuations in the horizontal 
plane and in the vertical direction.
This anisotropy has been investigated in many earlier studies,
in particular by analyzing the 
longitudinal and transverse energy spectra 
\cite{Brethouwer:07,Davidson_13,Rosenberg:15}. In the 
present work, we limit our discussion to  
the velocity components $u_\perp$ and $u_\|$, respectively in the 
horizontal and vertical directions  (perpendicular
and parallel to gravity).
The statistics involving $u_\perp$ are obtained by averaging the
two components in the x-y horizontal
plane (which is also the plane of rotation), whereas 
$u_\|$ simply corresponds to the vertical z-component.
An obvious manifestation of the anisotropy between the two directions
is in variances of $u_\perp$ and $u_\|$, as shown in Table~\ref{table:stat_v}.
It is evident that apart from the HIT run (case 0) and to a large extent
the rotation only run (case 8), the variances for
both components are very different.
To precisely quantify the anisotropy, we consider the ratio
$\xi^u = [ \langle u_\perp^2 \rangle/\langle u_\|^2 \rangle ]^{1/2}$,
also shown in Table~\ref{table:stat_v}.
Expectedly, this ratio is unity for case 0 and very close to unity for case 8.
However, for all runs with stratification, the ratio is always greater than 1.
For cases 1-3, the ratio increases slowly with decreasing $Fr$, whereas
a sharp increase occurs for cases 4-6 (and also case 7).
Clearly, this transition corresponds to the change in spectral slope
observed for the spectra in Fig.~\ref{fig:spctr}.
The runs with smaller $\xi^u$ correspond to spectra which show
$k^{-5/3}$ scaling, whereas runs with sharp increase of $\xi^u$
correspond to spectra with steeper Bolgiano-like scaling.
In these flows, the 
fluctuations of $u_\|$ are essentially suppressed when the imposed
RaS are strong.
We note that the values shown here are consistent with 
the results of some earlier runs~\cite{pouquet_18}.

\begin{table}[H]
\centering
    \begin{tabular}{l|c | cccccc |cc}
  \hline \hline
    case      &0    & 1     & 2      & 3     & 4   & 5   & 6 & 7 & 8    \\ 
\hline
    $Fr$   & $\infty$ & 0.168  & 0.114  & 0.086 & 0.069 & 0.063 & 0.045 & 0.030    & $\infty$  \\
    $Ro$   & $\infty$ & 0.840  & 0.570  & 0.430 & 0.345 & 0.315 & 0.225 & $\infty$ & 0.140     \\
\hline
    $\langle u_\perp^2 \rangle$  & 0.68 & 0.62    & 0.92    & 1.2      & 2.2    &  2.8    &  1.7    & 0.55      & 0.39      \\
    $\langle u_\|^2 \rangle$     & 0.67 & 0.28    & 0.17    & 0.097    & 0.022  &  0.019  &  0.005  & 0.0077    & 0.28      \\ 
    $\xi^u$                      & 1.01 & 1.48    & 2.33    & 3.52     & 10.0   &  12.0   &  18.4   & 8.4       & 1.18      \\ 
\hline
    $K_{u_\perp}$  & 2.9 &  2.9     & 2.9     & 2.9     & 3.1 &  3.4    & 3.05 & 2.5       & 3.1        \\
    $K_{u_\|}$     & 2.9 &  3.3     & 3.8     & 4.7     & 6.4 &  8.4    & 10.9* & 3.2       & 3.1        \\ 
\hline
    \end{tabular}
\caption{Variance and kurtosis of velocity components.
We use $\xi^u=\langle u_\perp^2 \rangle^{1/2}/\langle u_\|^2 \rangle^{1/2}$,
to measure the anisotropy, whereas $K_{u_\|}$ and $K_{u_\perp}$ are 
the kurtosis of $u_\|$ and $u_\perp$ respectively.
The skewness of these components (not shown) is consistent with $0$.
The value(s) of kurtosis marked with an asterisk correspond to relatively low accuracy.
}
\label{table:stat_v}
\end{table}

To further investigate the anisotropy,
we next consider higher order moments of the velocity components.
Since the probability density functions (PDFs)
of individual velocity
components are symmetric
(due to underlying symmetry of 
Boussinesq equations),
we note that the values of the third moments of the
velocity fluctuations are expected to be zero. 
Our results, not shown here, are compatible with this.
Thus, we consider the fourth-order moment,
through the kurtosis, defined as
$K_{u_i} = \langle u_i^4 \rangle/\langle u_i^2 \rangle^2 $.
It is known that the PDFs
of individual velocity
components in HIT are approximately 
Gaussian, 
which implies that the kurtosis is 
approximately $3$. 
Consistent with that, we find that the value of the
kurtosis is very close to $3$ in the case HIT,
(see Table~\ref{table:stat_v}).
In the case of a purely stratified
flow, however, it was found that the distribution of the velocity component 
parallel to the direction of stratification, $u_\|$, could be significantly
wider than Gaussian \cite{Rorai:14,Feraco:18}. 
The strong deviations from a Gaussian distribution
have been shown to originate from intermittent vertical drafts \cite{Feraco:18}.
On the other hand, the kurtosis of $u_\perp$ is always found to be 
approximately $3$.
Our values, as shown in Table~\ref{table:stat_v}, are 
consistent with these trends.
We find that 
$K_{u_\perp}$ is slightly smaller than $3$ at low 
stratification,
and slightly larger than $3$ for runs with strong stratification 
($Fr \lesssim 0.07$), with PDFs (not shown) being approximately Gaussian. 
In the vertical direction, $K_{u_\|}$ 
increases  very significantly beyond $3$
as strength of RaS increases.

It is worthwhile to compare the dependence of $K_{u_\|}$
in our RaS runs with the stratification-only runs 
reported in~\cite{Feraco:18} (see in particular 
their Fig.5).
While the sharp transition in $K_{u_\|}$ appears to be qualitatively 
similar, quantitatively the large values of 
$K_{u_\|}$ for RaS runs
occur at slightly lower $Fr$ values than in \cite{Feraco:18}.
For even smaller $Fr$ (not considered in this work), 
in a wave-dominated regime, 
one can expect that $K_{u_\|}$ for RaS runs
will again become 3 \cite{Feraco:18}, as it was the case
for the stratification-only run (case 7) discussed here.
Additional discussion of $K_{u_\|}$ will be presented in the following
section, in relation
with acceleration statistics.

\section{Acceleration statistics}
\label{sec:acceleration}

The acceleration experienced by a fluid element, 
defined by the rate of change of velocity in the Lagrangian frame, i.e., 
$\mathbf{a}^+ = d\mathbf{u}^+/dt$ (where $\mathbf{u}^+$ is defined in
Eq.~\eqref{eq:dxdt})
and resulting from the
balance of forces acting on it, is arguably
the simplest descriptor of its motion, as also directly reflected 
in the governing fluid equations.
In addition to its
fundamental importance in turbulence theory~\cite{TB_ARFM},
a key motivation 
to study acceleration comes 
from its central role in stochastic modeling 
of turbulent dispersion~\cite{wilson96}.
In the following sub-sections, we study 
different aspects related to acceleration, namely,
acceleration variance and kurtosis, 
the probability density functions and Lagrangian
autocorrelations and frequency spectra.
As already mentioned, the anisotropy of the flow makes it necessary 
to distinguish between
the horizontal and vertical components of acceleration, denoted by
$a^+_\perp$ and $a^+_\|$ respectively.
In this section, we demonstrate that the anisotropy properties of 
the acceleration components strongly differ from those of the velocity
discussed in the previous section.
For convenience, we will henceforth omit the superscript '+'
from our notation,
since our subsequent results only involve Lagrangian
quantities.

\subsection{Acceleration variance}
\label{subsec:acc_var}

\begin{table}
\centering
    \begin{tabular}{l|c | cccccc |cc}
  \hline \hline
 case & 0 & 1 & 2 & 3 & 4 & 5 & 6 & 7 & 8 \\
\hline
    $Fr$   & $\infty$ & 0.168  & 0.114  & 0.086 & 0.069 & 0.063 & 0.045 & 0.030    & $\infty$  \\
    $Ro$   & $\infty$ & 0.840  & 0.570  & 0.430 & 0.345 & 0.315 & 0.225 & $\infty$ & 0.140     \\
\hline
    $\langle a_\perp^2 \rangle$ & 16.4    & 6.9    & 6.3    & 5.4   & 5.4    & 9.3    & 6.21 & 0.48     & 2.1   \\
    $\langle a_\|^2 \rangle$    & 16.3    & 8.9    & 8.6    & 6.7   & 1.9    & 2.0    & 0.43 & 1.1      & 1.9   \\ 
    $\xi^a$                     & 1.01    & 0.88   & 0.85   & 0.90  & 1.67   & 2.11   & 3.80 & 0.66     & 1.05  \\ 
\hline
   $K_{a_\perp}$                & 19.3    & 29.8   & 37.1   & 46.9$^*$  & 6.52   & 7.50  & 14.3  & 4.38     & 22.8  \\
   $K_{a_\|}$                   & 19.6    & 22.3   & 27.2   & 40.5$^*$  & 17.9   & 11.8  & 16.2  & 3.27     & 35.6$^*$  \\ \hline
  { $\xi^u/\xi^a$  }                 & {1.0}    & {1.68}   & {2.74}   & {3.9}  & {6.0}   & {5.7}  & {4.8}  & {12.7}     & {1.1 } \\ \hline
    \end{tabular}
\caption{Variance and kurtosis of acceleration components, perpendicular and parallel
to the axis of rotation/stratification.
$\xi^a = [\langle a_\perp^2 \rangle/\langle a_\|^2 \rangle ]^{1/2}$ measures 
the anisotropy, whereas the kurtosis, 
$K_{a_\perp}$ and $K_{a_\|}$, characterizes the 
extent of the PDFs (see Fig.~\ref{fig:acc_PDF}).
The underlying symmetry of the problem imposes that the skewness (not shown) 
is zero.
The values of kurtosis marked with an asterisk correspond to relatively low accuracy.
}
\label{table:acc}
\end{table}

Extending the earlier analysis of the Eulerian velocity field,
we first examine the anisotropy in acceleration by considering the
second and fourth order moments, namely, the variance
and kurtosis respectively.
The moments are listed in Table~\ref{table:acc}. 
Due to the underlying symmetry of the Boussinesq equations, 
the third and all other odd moments of acceleration components
are zero
(similar to the velocity components). 
As can been seen from cases 1-6, the properties of $a_\perp$ and $a_\|$
show striking differences.
While the variances of $a_\|$ rapidly decrease 
with decreasing $Ro$ (and $Fr$), the variances of 
$a_\perp$ do not change that much.
To better understand this behavior, we consider the
anisotropy ratio  
$\xi^a = [ \langle a_\perp^2\rangle/\langle a_\|^2 \rangle ]^{1/2}$.
For the HIT run (case 0), the ratio, as expected, is equal to 1, 
whereas for rot-strat cases two separate behaviors are visible.

The first corresponds to 
cases 1-3 with relatively weak rotation and stratification (RaS),
where the ratio $\xi^a$ is approximately constant, and slightly 
smaller than 1.
The ratio being close to unity suggests that the effect of anisotropy
on small scales is only minor. 
Furthermore, a simple explanation for the ratio slightly smaller
than unity can be provided based on the knowledge that
the relative strength of stratification is significantly
stronger compared to that of rotation (since $N/f=5$).
For that reason, the acceleration
variance in the vertical direction is likely to be 
slightly more enhanced (due to stratification) than
that in the horizontal direction (due to rotation).
On the other hand, runs 4-6 correspond to significantly 
stronger RaS, which leads to
a different behavior, with $\xi^a > 1$ and further increasing with 
decreasing $Fr$ (and $Ro$).  
Since runs 4-6 correspond to $R_{IB}<1$, they are in a regime
where waves play a stronger role, compared to (turbulent) eddies.
With strong stratification, the motion in the
vertical direction is strongly suppressed, as also
reflected in $\xi^u$ values (see Table~\ref{table:stat_v}).
With relative weaker effect of rotation, the acceleration
in the horizontal direction is still affected by turbulent eddies,
resulting in $\xi^a$ to become greater than 1. 
Thereafter, the effect becomes even more pronounced with further
increase in strength of RaS.

It is instructive to compare the acceleration variances  
for case 5 (with strong RaS), to 
case 7, with exactly the same stratification, $N$, but no rotation 
and to case 8, with the same rotation, $f$, but no stratification. 
The variances of $a_\perp$ and $a_\|$ are stronger in the presence of 
both RaS, than with either of the two effects taken separately. 
The anisotropy ratio $\xi^a$ also shows a very striking difference, when 
comparing runs 5, 7 and 8. For rotation-only (run 8), 
we observe $\xi^a \approx 1$, implying little to no small-scale anisotropy. This can be 
viewed as consistent with
the observation that the flow spectrum for run 8 is overall comparable to 
the Kolmogorov
spectrum (see Fig.~\ref{fig:spctr}a), so the flow is dominated by eddies.
On the other hand, the anisotropy 
ratio is $\xi^a < 1$ ($0.66$) for run 7, implying stronger acceleration 
variances in the
vertical than in the horizontal direction, but an opposite situation
in the presence of both RaS: $\xi^a > 1$ ($2.1$).
This difference can be traced back to the much larger
value of $\langle a_\perp^2 \rangle$ in the presence of rotation, a
consequence of the enhancement of the horizontal component of acceleration
due to the Coriolis force.

In comparing the anisotropy of acceleration components
with that of velocity (see Table~\ref{table:stat_v}),
we find that $\xi_a$ is always smaller 
than $\xi_u$, and
additionally very close to unity when $R_{IB}>1$.  
This can be explained by realizing that velocity is sensitive
to the large scales of the flow and is directly influenced
by the imposed anisotropy. On the other hand,
acceleration mostly samples the small scales,
and is not likely sensitive to the imposed anisotropy (at large scales), 
provided the scale separation between the large and small
scales is sufficiently wide -- consistent
with the notion of local isotropy as dictated
by Kolmogorov's theory \cite{MY.I}.
In the current scenario, this is realized
when $R_{IB}>1$, resulting in 
$\xi^a$ being nearly unity
(and also a $k^{-5/3}$ range
in the intermediate $k$ range as
demonstrated earlier).
Whereas for $R_{IB}<1$, local isotropy is
clearly violated, but still the anisotropy
of acceleration is significantly weaker than that of velocity.

Additional insight can be obtained by
considering the ratio $\xi^u/\xi^a$, also shown in Table~\ref{table:acc}.
For HIT, this ratio is unity, but with increasing
strength of RaS, we observe $\xi_u/\xi_a$ increases, peaking
at around 6 for case 4, and thereafter decreasing slightly
with further increase in strength of RaS.
A limiting value can be obtained in the case where
RaS is very strong, such that
the flow
is dominated by linear processes
(with turbulent eddies playing a very weak role).
This would lead to $\xi^u/\xi^a \approx N/f$, by assuming
that $a_\perp \sim u_\perp f$ and $a_\| \sim u_\| N$.
The value for case 6, with $\xi^u/\xi^a = 18.4/3.8 \approx 5$,
appears to be consistent with this  consideration,
though it remains to be further tested for different $N/f$ values.

\begin{figure}
\centering
\includegraphics[width=8.0cm]{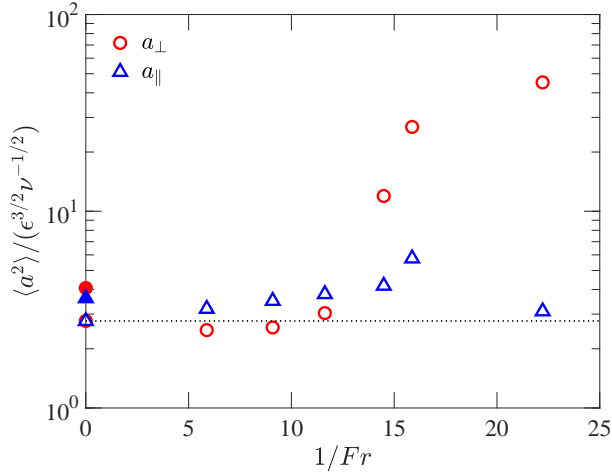}
\caption{
Kolmogorov-scaled acceleration variance, 
as defined by Eq.~\ref{eq:a0}, plotted as a function of 
$1/Fr$. The parallel (vertical)
and perpendicular (horizontal) components
of accelerations 
are shown in blue triangles and red circles respectively.
The HIT run corresponds to $1/Fr=0$. In addition the rotation-only 
run also corresponds to $1/Fr=0$, but is shown in solid symbols.
}
\label{fig:a_0}
\end{figure}

Given that the statistics of acceleration 
appear to conform with local isotropy for $R_{IB}>1$,
we consider
the following relation from classical turbulence:
\begin{equation}
\langle a^2 \rangle = a_0 \langle \epsilon \rangle^{3/2} \nu^{-1/2} 
\label{eq:a0}
\end{equation}
which results from the application of 
Kolmogorov's similarity theory to acceleration variance \cite{MY.I}.
Here, $a_0$ is a dimensionless constant, which is plausibly universal
for HIT.
Empirical evidence based on studies of isotropic turbulence
has shown that $a_0$ increases slowly 
with Reynolds numbers~\cite{Sawford03}, which can be possibly 
viewed as a manifestation of small-scale intermittency, 
unaccounted for in
Kolmogorov's 1941 theory. 
In contrast, for RaS turbulence, one may expect
additional deviations because of the imposed anisotropy, especially 
if $Re$ is not very large. 

Fig.~\ref{fig:a_0} shows the values of the $a_0^{\perp}$ and 
$a_0^{\|}$ as a function of $Fr$ for cases 1 through 6, together with the
value obtained for case 0, in the absence of RaS.
The values of $a_0^{\|}$ and $a_0^{\perp}$ 
do not vary very much for 
$1/Fr \gtrsim 12.5$, 
and sharply increase, 
especially in the perpendicular direction,
at large values of $1/Fr$. In particular, the value of 
$a_0^{\perp}$ at $1/Fr \approx 16$
exceeds that in the HIT case by a factor larger than  $\sim 10$. 
This points to a strong difference of the small-scale properties
between runs 4-6 and the HIT run (case 0).
In comparison, and 
despite the larger value of $N$ 
($N/f = 5$ for runs 1-6), the value of $a_0^{\|}$ remains 
comparable to that obtained in the HIT case. 
In the purely stratified case (run 7), the value of $a_0^{\|}$ is 
slightly reduced compared to case 5 in the presence of both RaS
($a_0^{\|} \approx 3.1$). On the other hand, the value of 
$a_0^{\perp} \approx 1.4$ for case 7 is reduced 
by over an order of magnitude compared to case 5. 
For the rotation-only run, $a_0$ is not very different compared to the
HIT case, suggesting that the small scales are still dominated by turbulence,
even though the large scales are highly anisotropic.
This once again demonstrates that acceleration becomes progressively dominated, as 
$Ro$ and $Fr$ decrease, by the frequencies associated with rotation/buoyancy.
As a consequence, the values of $a_0$ shown in Fig.~\ref{fig:a_0} do not
reflect only the small-scale structure of the flow, but also its global
properties.

\subsection{Acceleration PDFs}
\label{subsec:PDF_acc}

\begin{figure}
\begin{center}
\includegraphics[width=8.0cm]{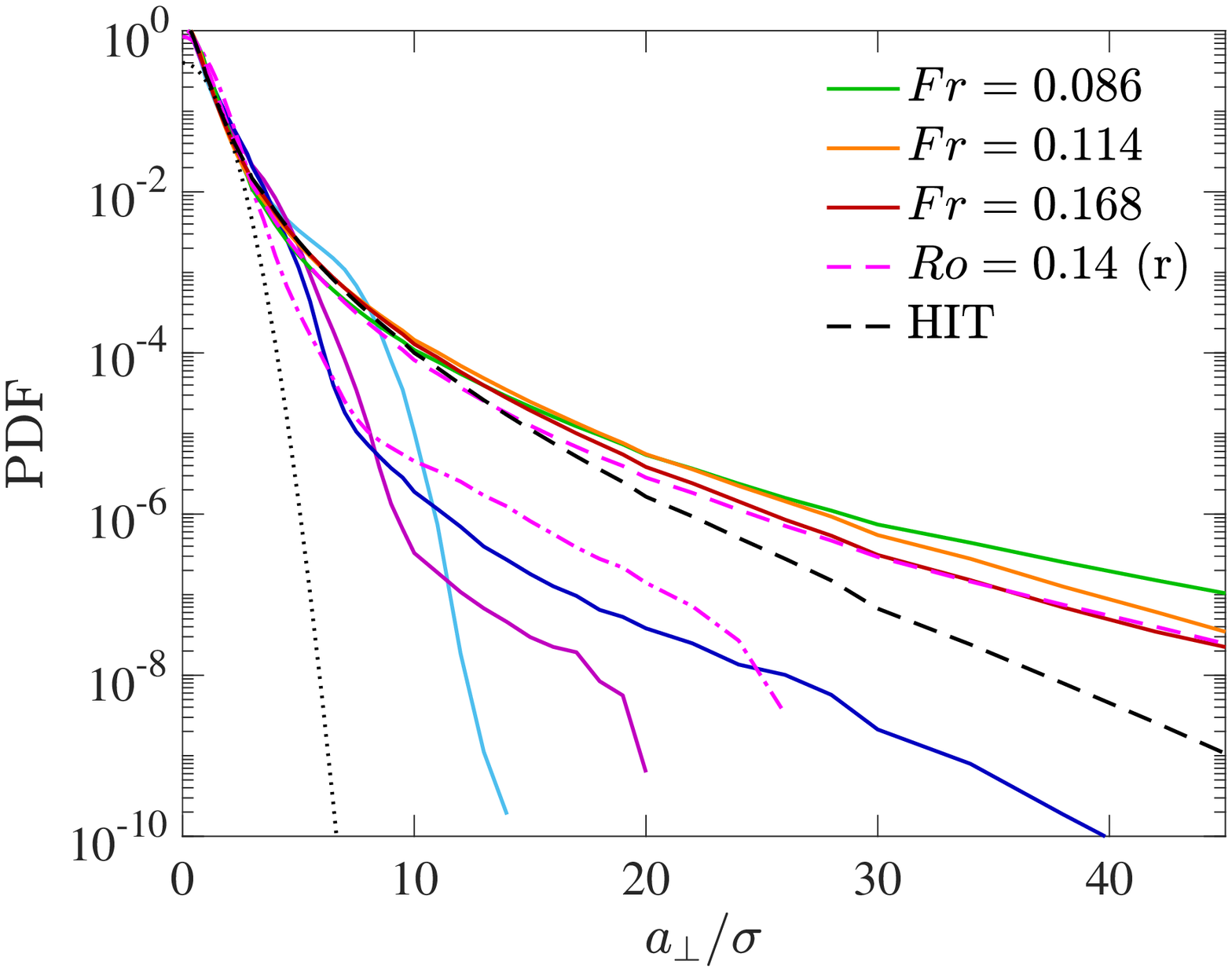}
\includegraphics[width=8.0cm]{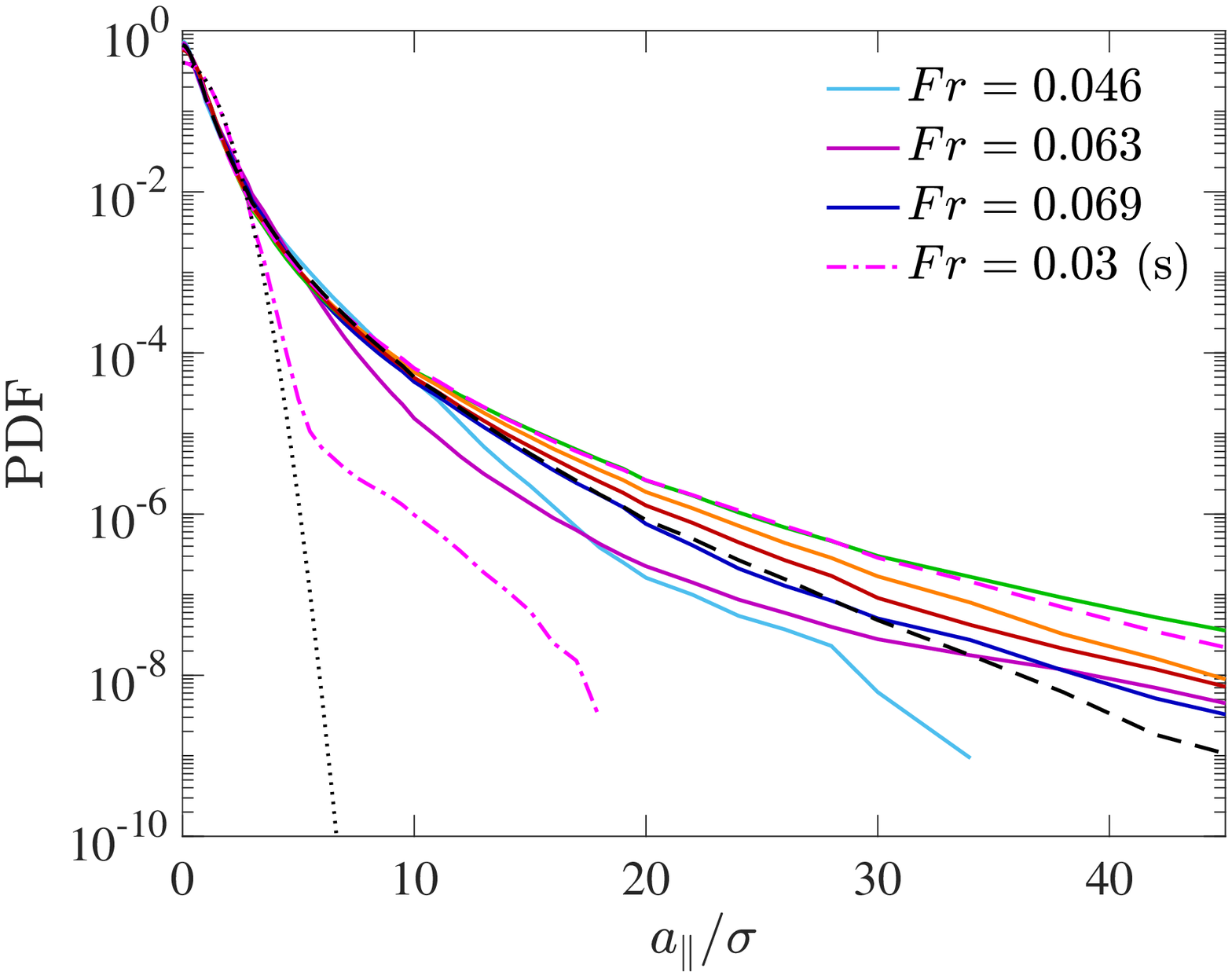} \\
\caption{
Standardized probability density functions (PDFs) of 
(a) the perpendicular (horizontal) and 
(b) the parallel (vertical)
components of acceleration, i.e., $a_\perp$ and $a_\|$ respectively. 
$\sigma$ denotes the corresponding standard deviation.
The dotted line shows the standardized Gaussian distribution for comparison. 
The legend is split over two panels, but applies to each panel individually. 
}
\label{fig:acc_PDF}
\end{center}
\begin{picture}(0,0)(0,0)
\thicklines
%\put(-175,440){$(a)$}
%\put(60,440){$(b)$}
\put(-165,260){$(a)$}
\put(70,260){$(b)$}
\end{picture}
\end{figure}

The analysis of acceleration variance can be generalized
by considering the probability density functions (PDFs) of 
acceleration components. 
In HIT,
similar to Eulerian velocity gradients, 
acceleration is also characterized by extremely
large fluctuations reflected in broad tails of the PDF,
which further broaden with increasing Reynolds number
\cite{TB_ARFM, BPBY2019}.
In this subsection, we study the effect of imposed RaS
on the PDFs of acceleration, especially
focusing on the long tails.  
Once again, we separate out the contributions $a_\|$ and $a_\perp$.

Fig.~\ref{fig:acc_PDF}a and b respectively show the standardized PDFs of $a_\perp$
and $a_\|$ for all the runs listed in 
Table~\ref{table:acc}.
Two very distinct behaviors can be observed once again.
At moderate RaS (and for $R_{IB}>1$) 
the PDFs of acceleration exhibit very broad tails, and in this sense, they
differ only quantitatively from those obtained in the HIT case. 
This property is also reflected by the values of the kurtosis of the distributions,
shown in Table~\ref{table:acc}, which are all $\gtrsim 20$, 
even larger than the 
values obtained for the HIT case. 
The run corresponding to pure rotation also exhibits
very high kurtosis of the acceleration. Note that the accuracy of 
the estimates of kurtosis indicated in the table
are of the order of $10\%$ or less, except for values marked with an asterisk,
which exhibited very large fluctuations, leading to higher error bars
for runs 3 and 8
(of about $20-30\%$).

In contrast for the runs with strong RaS (with $R_{IB}<1$), 
the tails of the PDFs are significantly suppressed. As shown
in Table~\ref{table:acc}, the corresponding kurtosis values are also
quite small, although still larger than the Gaussian value of 3.
This is particularly clear for the horizontal component $a_\perp$ of
acceleration, see Fig.~\ref{fig:acc_PDF}a, and to a lesser extent, for the
vertical component $a_\|$ (Fig.~\ref{fig:acc_PDF}b). 
The tendency of the PDFs to become narrower in runs with 
strong RaS appears to be a
consequence of the strong stratification, as the PDFs of acceleration
in the run with pure rotation are only weakly modified.
In regimes dominated by waves, weak turbulence 
theory suggests 
that the velocity and acceleration should be 
close to Gaussian \cite{Nazarenko:2011}.
However, we observe that the 
kurtosis of acceleration components are significantly larger than 
the corresponding Gaussian value of $3$ (with a stronger
kurtosis for the vertical, than for the horizontal component of acceleration).
This unexpected phenomenon can also be associated with the observation of bursts of
vertical velocity, manifested by the large values of $K_{u_\|}$ listed
in Table~\ref{table:stat_v}, particularly for runs 4-6. These bursts
are related to the phenomenon observed in~\cite{Feraco:18} in the case
of purely stratified flows, in comparable domains of parameters. We expect
that the mechanisms are similar in the presence of a 
relatively weak rotation 
($N/f = 5$), as it is the case in the present study.

\subsection{Acceleration autocorrelation and frequency spectra}
\label{subsec:Corr_acc}

Essential information on the motion of tracers can be obtained from 
the autocorrelation function of quantities fluctuating along Lagrangian
trajectories. For example,
the autocorrelation function can be used to determine relevant time scales
and to form the frequency spectrum via Fourier transform \cite{Yeung89}.
Since our analysis is concerned with statistically stationary
signals, the autocorrelation only depends on the chosen time lag.
The autocorrelation $\rho^a(\tau)$ is defined as
\begin{align}
C^a(\tau) = \langle a(t+\tau) a (t) \rangle \ &, \ \   
 \rho^a(\tau) = C^a(\tau)/C^a(0) 
\label{eq:def_C_a}
\end{align}
where $\tau$ is the time lag and $C^a$ is the autocovariance. 
Statistical stationarity implies $\rho^a(-\tau) = \rho^a(\tau)$
and also imposes that the integral of the acceleration autocorrelation
must be zero, i.e., $\int_0^\infty \rho^a(\tau) d\tau=0$ \cite{Yeung89}.
The frequency spectrum of acceleration $E^a$, which is the Fourier transform of 
$C^a(\tau)$ can be defined as 
\begin{align}
E^a(\omega) = \int_0^\infty  C^a(\tau) \exp^{-i\omega\tau} d\tau 
\end{align}
By definition, the integral of the spectrum gives the variance of the signal.
Yet again, we consider $a_\perp$ and  $a_\|$ separately.

\paragraph*{Perpendicular component:}
%\label{subsubsec:perpendicular}

\begin{figure}
\begin{center}
\includegraphics[width=7.9cm]{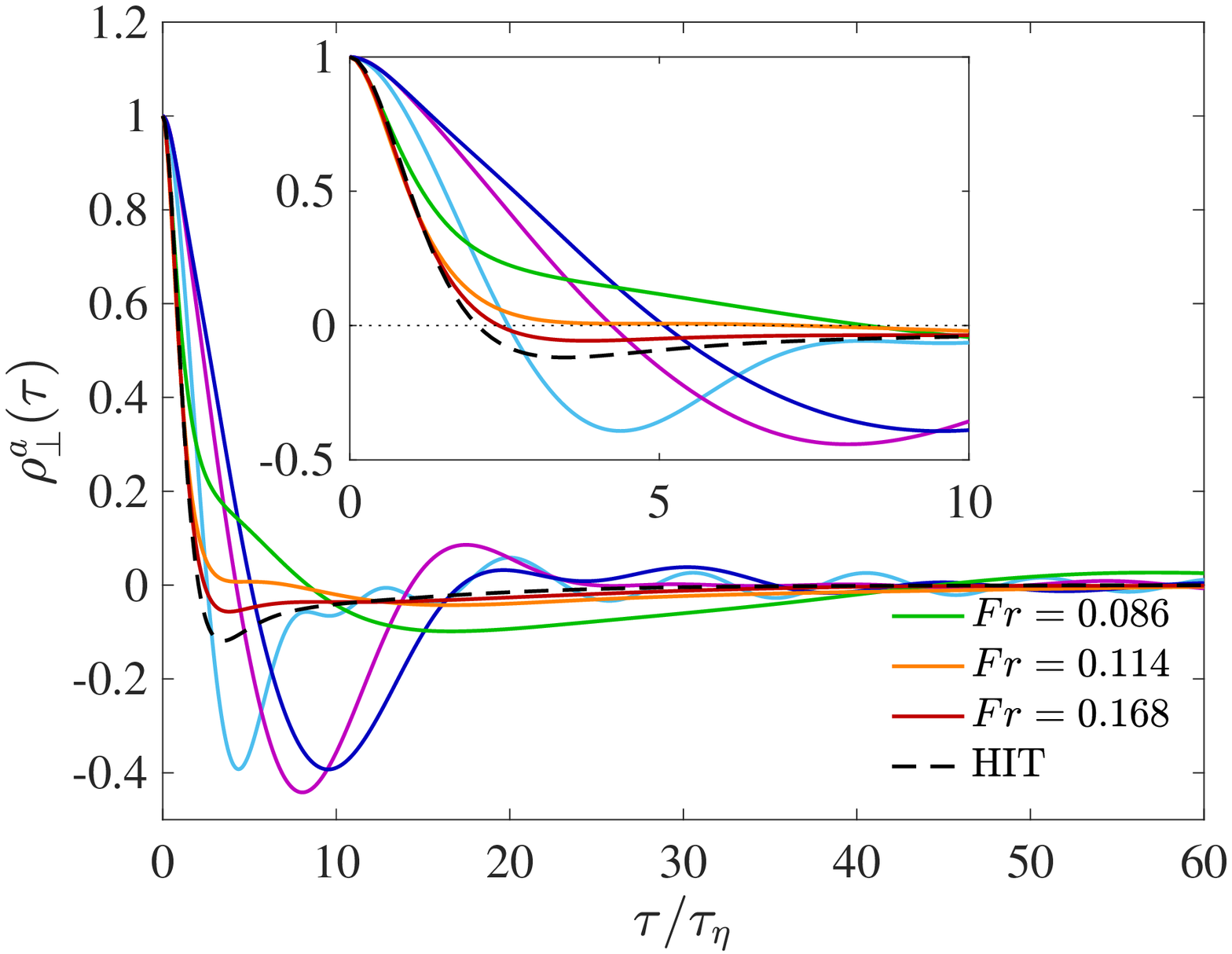}
\includegraphics[width=7.9cm]{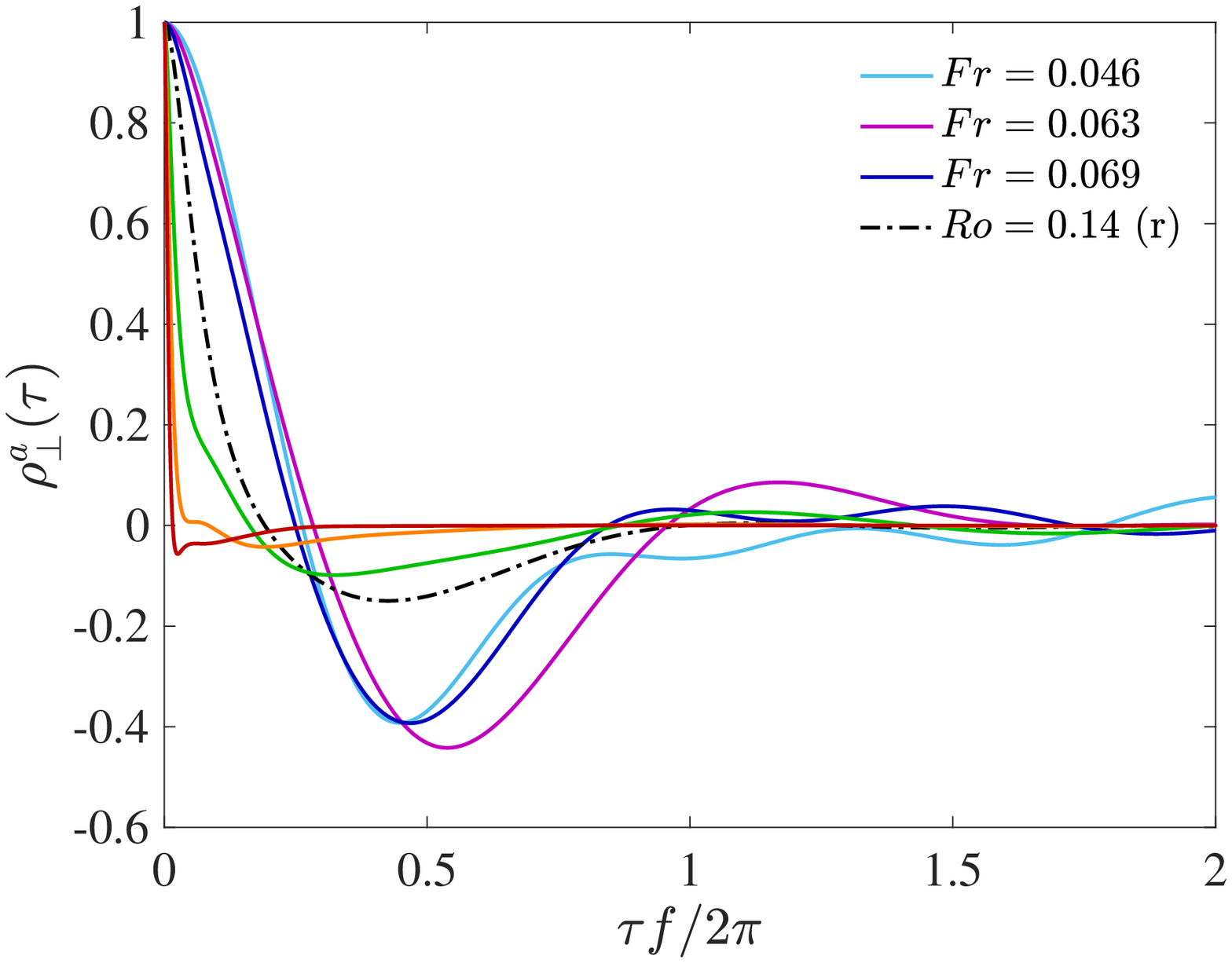}
\caption{
Lagrangian autocorrelation of the perpendicular (horizontal) component of
acceleration $\rho^a_\perp(\tau)$ against the time lag $\tau$ 
normalized by the Kolmogorov time scale
$\ktime$ in (a) and by the rotation period $2\pi/f$ in (b). 
Inset in (a) shows a zoomed view at small time lag.
The legend is split over two panels, but applies to each panel individually. 
}
\label{fig:aut_a_perp}
\end{center}
\begin{picture}(0,0)(0,0)
\thicklines
\put(-35,230){$(a)$}
\put(60,230){$(b)$}
\end{picture}
\end{figure}

Fig.~\ref{fig:aut_a_perp} shows the autororrelation function
$\rho^a_\perp(\tau)$ for all the runs with RaS.
In Fig.~\ref{fig:aut_a_perp}a, the time lag is normalized with $\tau_\eta$, 
the Kolmogorov
time scale characterizing the small-scale motion.
We also include
the HIT case (dashed line) and the pure rotation case (dash-dotted line) as a reference.  
%For HIT, it is well known that the acceleration autocorrelation decays
For HIT, the acceleration autocorrelation is known to decay
rapidly, becoming zero at $\tau \approx 2\tau_\eta$
and thereafter becoming negative and slowly approaching zero again
at very large time lags \cite{yeung:07} -- as readily seen in
Fig.~\ref{fig:aut_a_perp}a.
Interestingly, we observe that the runs with relatively weak RaS,
i.e., $Fr \ge 0.114$, and also the run
with only rotation, behave very similarly to the HIT case
(see inset of Fig.~\ref{fig:aut_a_perp}a), albeit with minor 
variation in the zero-crossing point. 
For the rot-strat run with $Fr=0.086$, the autocorrelation function
shows a similar behavior up to very small time lags, but thereafter
deviates with the zero-crossing extending to $\tau \approx 8\tau_\eta$.
The large time behavior is markedly different with very low frequency
oscillations. 
On the other hand, for runs with strong RaS,
i.e., $Fr \le 0.07$, a very different behavior is observed
right from small time lags. The autocorrelation shows strong oscillations 
which are eventually damped out at large time lags.
Furthermore, the zero-crossing point also is strongly dependent on
$Fr$, moving to smaller time lags with decreasing $Fr$.

Plotting $\rho^a_\perp$ as 
a function of $\tau f /2 \pi$, see Fig.~\ref{fig:aut_a_perp}b, shows that
the period of the damped oscillations is close to $2 \pi/f$ 
(with small but significant deviation).
For comparison, we also show in Fig.~\ref{fig:aut_a_perp}b the
function $\rho^a_\perp$ for the case of a purely rotating flow. Somewhat 
surprisingly, the tendency to oscillate at a frequency $\sim f$ is not 
as strong as in the case where both RaS are present.
A similar trend was observed in~\cite{DelCastello+:11a}. 
We notice, however, that the stronger stratification ($N=5f$ in runs 1-6) 
implies the presence of faster waves than in a purely rotating flow
at the same $Ro$ number, involving in particular high values of 
$k_\perp$, since the dispersion law for inertia-gravity waves can be 
written as  $k^2\omega_{IG}^2=N^2k_\perp^2+f^2k_\parallel$ \cite{pedlosky}.

%and 
%that such waves are expected to dominate at small scales, 
%for example in the process of recovering isotropy beyond the 
%Ozmidov scale  \cite{sagaut_08,marino_15w, herbert_16,pouquet_19p}.

\begin{figure}
\begin{center}
\includegraphics[width=7.9cm]{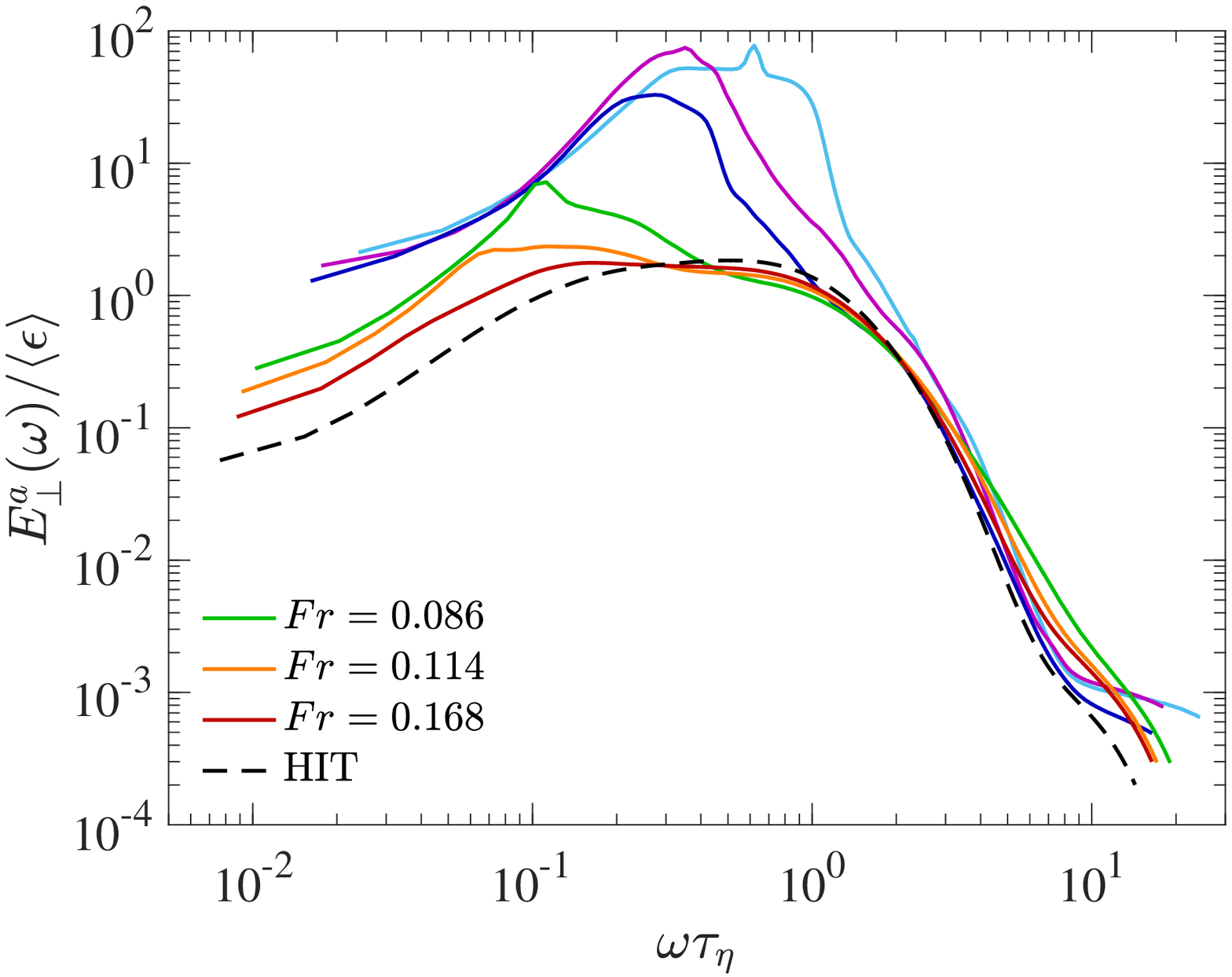}
\includegraphics[width=7.9cm]{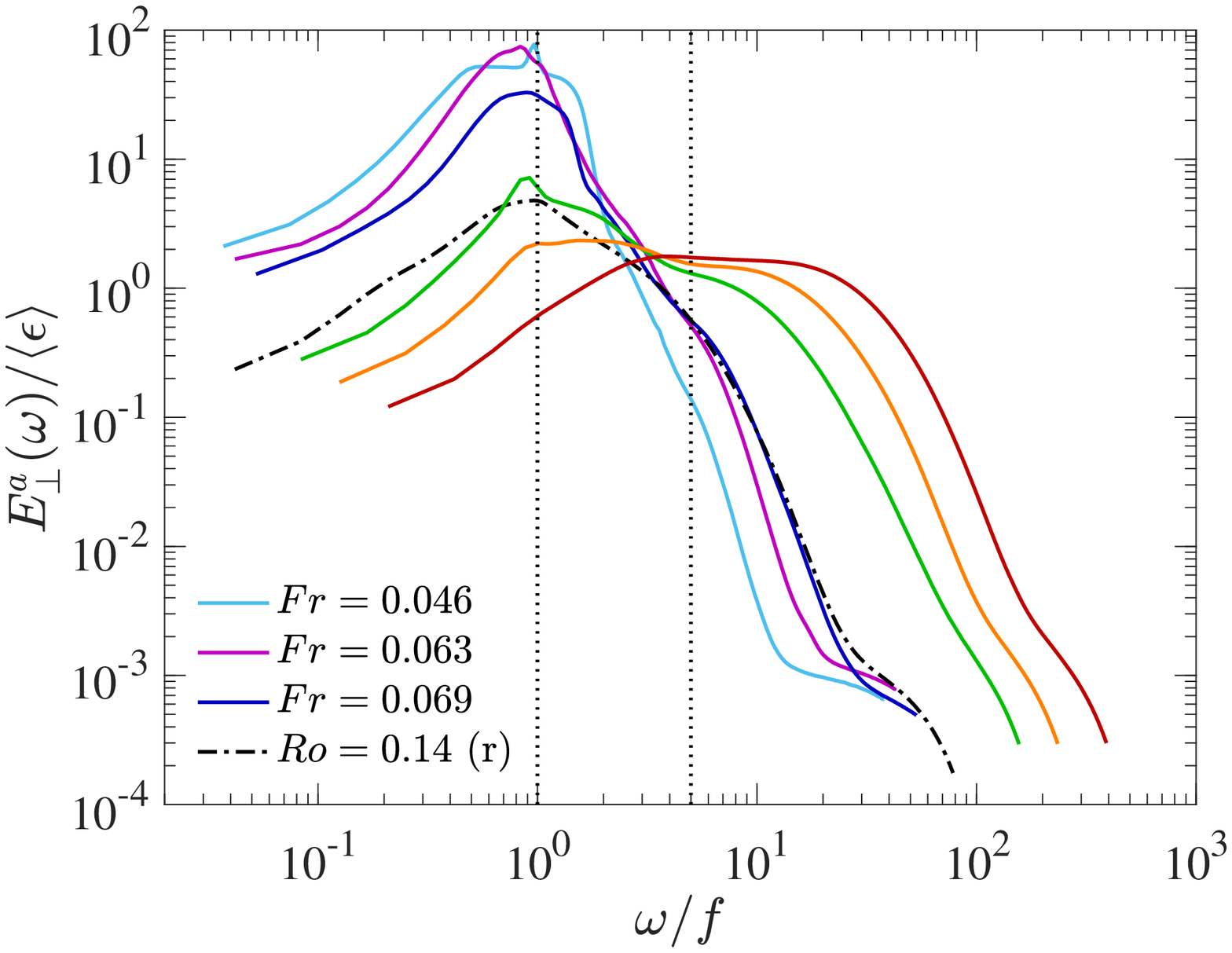} 
\caption{
Lagrangian frequency spectrum of the perpendicular (horizontal) component of
acceleration $E^a_\perp(\omega)$ against the frequency $\omega$ normalized by the 
Kolmogorov time scale
$\ktime$ in (a) and by the rotation frequency $f$ in (b).
The spectra are non-dimensionalized using the mean kinetic energy dissipation rate 
$\langle \epsilon \rangle$.
Vertical dotted lines in (b) correspond to frequencies $\omega=f$ and 
$N$ (note $N=5f$). 
The legend is split over two panels, but applies to each panel individually. 
}
\label{fig:spctr_a_perp}
\end{center}
\begin{picture}(0,0)(0,0)
\thicklines
\put(-35,215){$(a)$}
\put(190,215){$(b)$}
\end{picture}
\end{figure}

To better identify the frequencies 
present in the autocorrelation function, it is apt 
to take its Fourier transform to obtain the frequency spectrum.
Fig.~\ref{fig:spctr_a_perp} shows the spectrum $E^a_\perp$.
The spectra have been made dimensionless by dividing by the mean kinetic energy dissipation 
rate $\langle \epsilon \rangle$.
In Fig.~\ref{fig:spctr_a_perp}a, the frequency is normalized by 
$\tau_\eta$, whereas in Fig.~\ref{fig:spctr_a_perp}b, we use the 
rotation frequency $f$. Note that the runs shown in 
Fig.~\ref{fig:spctr_a_perp}a and b correspond to the same runs shown
respectively in Fig.~\ref{fig:aut_a_perp}a and b.
For the HIT run in Fig.~\ref{fig:spctr_a_perp}a, 
Kolmogorov scaling would imply that the 
spectrum $E^a$ is constant in the inertial range (defined as 
$ \tau_\eta/T_E \lesssim \omega \tau_\eta \lesssim 1$).
%where $T_E = L/U$ is the eddy turnover time). 
While the validity of Kolmogorov's similarity hypotheses to 
Lagrangian statistics is debatable, 
DNS studies suggest that the above scaling law, i.e., $E^a \sim \epsav$,
still may be approximately satisfied \cite{Sawford:11}.
Consistent with this, we find that for our HIT run, the spectrum
$E^a_\perp(\omega)$ is approximately flat over a limited range of values of 
$\omega$. 

With addition of weak RaS, the 
changes in the structure of the spectrum 
appear to be relatively minor.
In particular, for runs with $Fr \ge 0.086$, the differences
in spectra, relative to the HIT case, only become significant
for $\omega\tau_\eta \lesssim 0.2$.
For the particular case of $Fr = 0.086$, a small peak emerges
around $\omega\tau_\eta \approx 0.1$. 
With increasing strength of RaS,
the differences become more pronounced, with all the cases
for $Fr \le 0.086$ now showing a prominent peak (which is relatively broad).
This observation is consistent 
with the behavior of the autocorrelation function 
at small times, as observed in Fig.~\ref{fig:aut_a_perp}. 
In addition, 
the increase in values of $E^a_\perp/\langle \epsilon \rangle$ also 
corresponds to the strong increase in the value of $a_0$
for small $Fr$ cases, as shown in Fig.~\ref{fig:a_0}
(for the perp. data points).
Interestingly, the decay of spectra for
all cases are somewhat similar at very large frequencies
(and not very different from the HIT case).  
This possibly confirms that the role of turbulent eddies
in the horizontal direction is still relevant 
as compared to the vertical direction, at least in the 
parameter range covered in this work
(note once again the fact that $N/f=5$, plays
an important role, since it renders the effect of rotation
relatively weaker than that of stratification).

Fig.~\ref{fig:spctr_a_perp}b shows the spectra as a function of
$\omega/f$. 
As expected, the peaks observed at low frequencies
in Fig.~\ref{fig:spctr_a_perp}a, are all centered around $\omega/f \approx 1$.
For the runs with strong RaS, the frequency 
corresponding to stratification, i.e., $\omega/f = N/f = 5$,
does not seem to play any particularly important role.

\paragraph*{Parallel component:}

\begin{figure}
\begin{center}
\includegraphics[width=7.9cm]{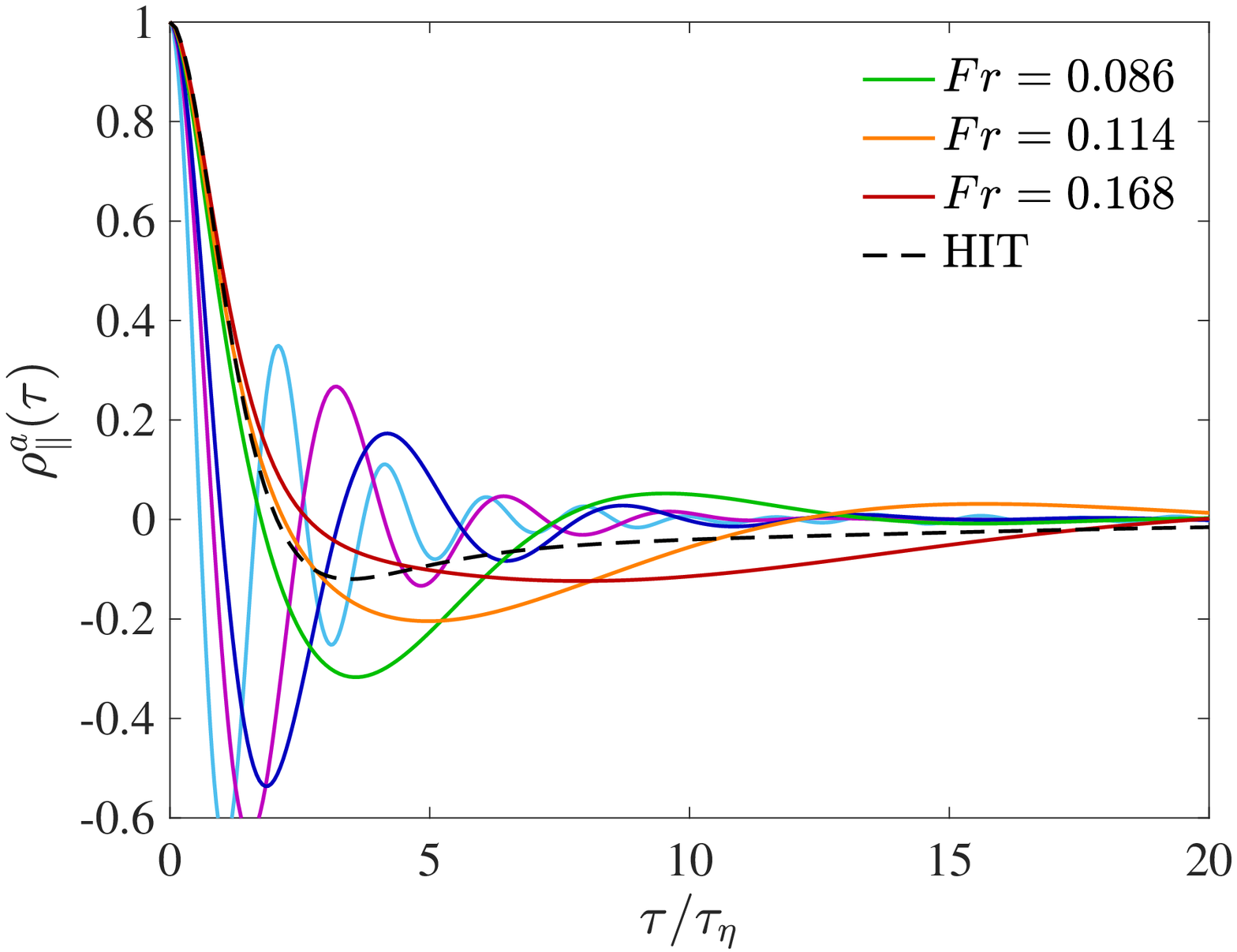}
\includegraphics[width=7.9cm]{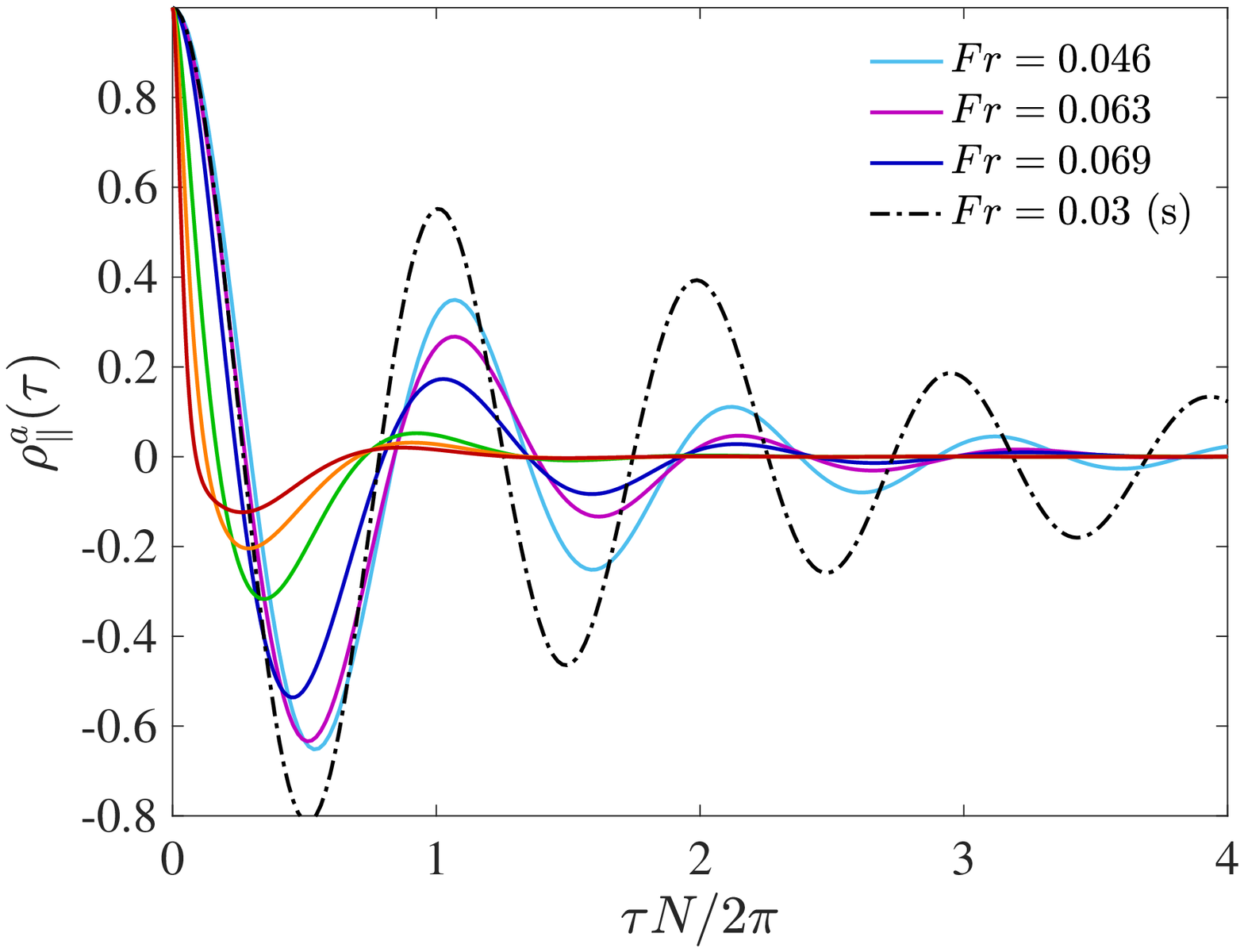}
\caption{
Lagrangian autocorrelation of the parallel  (vertical) component of
acceleration $\rho^a_\| (\tau)$ against the time lag $\tau$ normalized by the Kolmogorov time scale
$\ktime$ in (a) and by the Brunt-V\"ais\"al\"a period $2\pi/N$ in (b). 
The legend is split over two panels, but applies to each panel individually. 
}
\label{fig:aut_a_par}
\end{center}
\begin{picture}(0,0)(0,0)
\thicklines
\put(-175,215){$(a)$}
\put(50,215){$(b)$}
\end{picture}
\end{figure}

While the role of rotation was dominant in the 
perpendicular direction, we can in turn expect
stratification to dominate in the parallel direction.
In Fig.~\ref{fig:aut_a_par}a and b, we show the autocorrelation
in the parallel direction $\rho^a_\|$ 
as a function of 
$\tau/\tau_\eta$  and of $\tau (N/2 \pi)$ respectively. As it was the
case for the perpendicular component of acceleration, the deviation of the
autocorrelation function from the reference 
homogeneous isotropic case, shown as a dashed line, increases gradually 
when $Fr$ decreases. However, contrary to $\rho^a_\perp$, 
$\rho^a_\|$ tends to decay significantly faster than in the HIT case when 
$Fr$ diminishes. 
In addition, for all cases, overshooting oscillations 
are clearly visible, with a  particularly strong amplitude
for cases with $Fr \le0.086$.
While the zero-crossing of 
$\rho^a_\|$ for cases with weak RaS
still appears to be close to $\tau \approx 2\tau_\eta$ 
(with minor deviations), for cases with 
strong RaS, the zero-crossing
shifts to even smaller time lags of $\tau \approx 0.5\tau_\eta$.
This is in contrast to $\rho^a_\perp$, where the zero-crossing
first shifted to higher time lags, and then moved to smaller
time lags with decreasing $Fr$.
Fig.~\ref{fig:aut_a_par}b clearly
reveals that the period of the oscillations is equal to the
Brunt-V\"ais\"al\"a period, $2 \pi/N$.
This strongly suggests that the motion 
in the parallel direction is dominated by stratification, and that turbulence
plays a much smaller role for the vertical motion
(in comparison to horizontal motion, where even at strongest rotation
rate, the role of turbulence still could not be ignored).
For comparison, the autocorrelation
function $\rho^a_\|$ is shown in Fig.~\ref{fig:aut_a_par}b 
for the run with a strong stratification, 
$Fr = 0.03$. The tendency to oscillate is even stronger in this run, an 
effect amplified by the low value of $Fr$ (and $R_{IB}$) in this case (see
Table~\ref{table:stat_v}).

\begin{figure}%[H]
\begin{center}
\includegraphics[width=7.9cm]{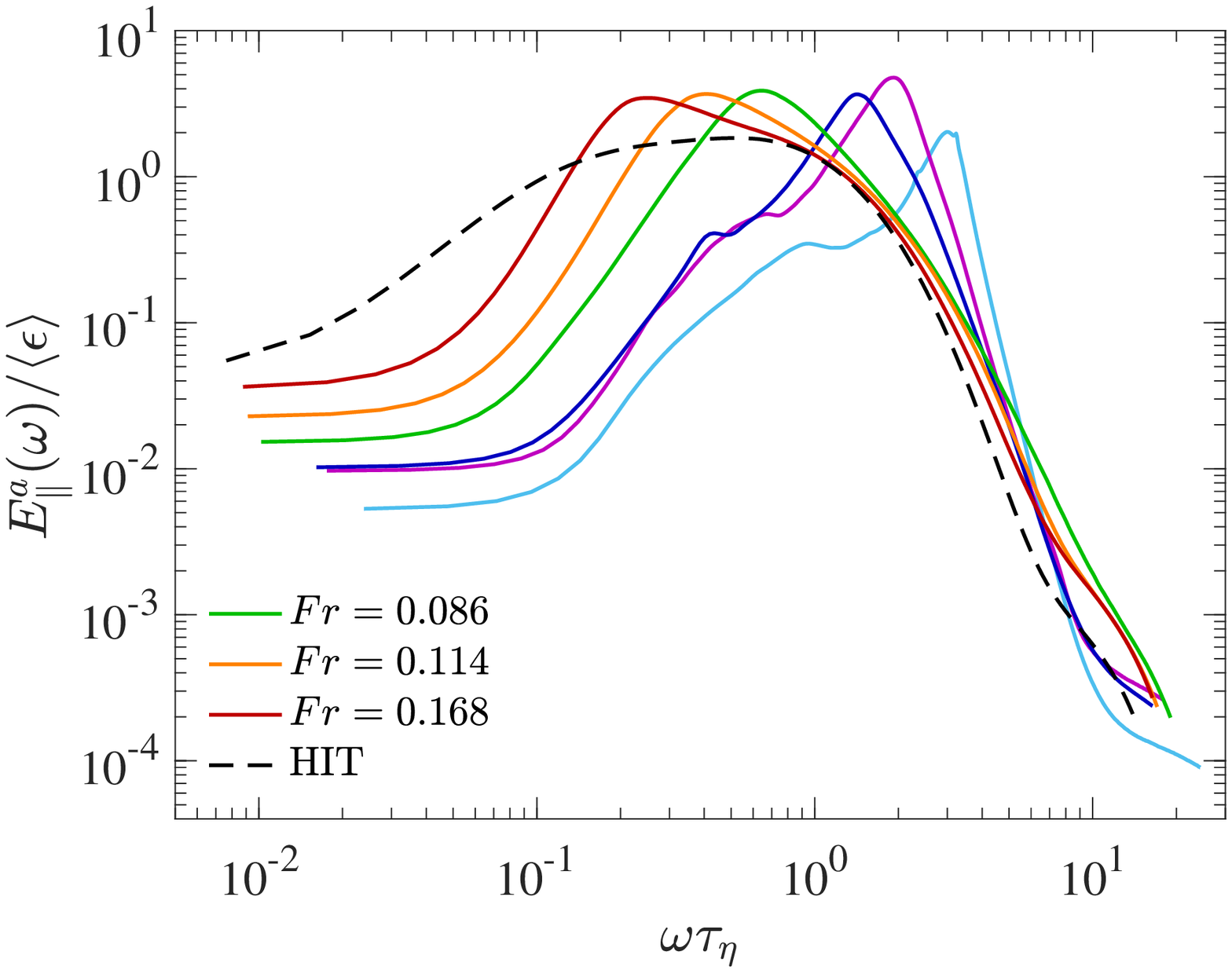}
\includegraphics[width=7.9cm]{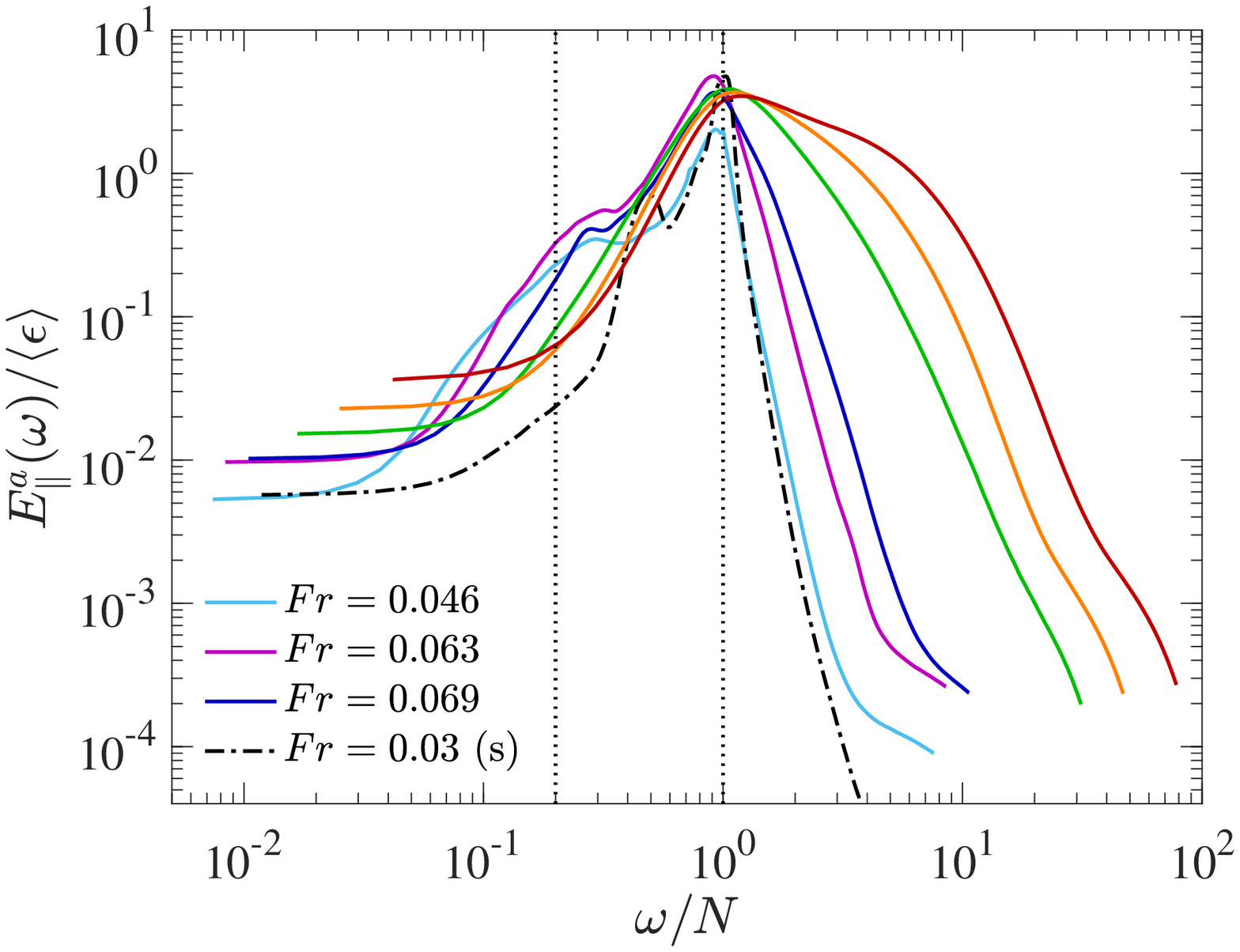}
\caption{
Lagrangian frequency spectrum of the parallel (vertical) component of
acceleration $E^a_\|(\omega)$ against the frequency $\omega$ normalized by the Kolmogorov
time scale
$\ktime$ in (a) and by the Brunt-V\"ais\"al\"a frequency $N$ in (b).
The spectra are non-dimensionalized using the mean kinetic energy dissipation rate
$\langle \epsilon \rangle$.
Vertical dotted lines in (b) correspond to frequencies $\omega=f$ and $N$ (with $N=5f$).
The legend is split over two panels, but applies to each panel individually. 
}
\label{fig:spctr_a_par}
\end{center}
\begin{picture}(0,0)(0,0)
\thicklines
\put(-35,225){$(a)$}
\put(190,225){$(b)$}
\end{picture}
\end{figure}

The corresponding frequency spectra
$E^a_\|(\omega)$ are plotted as a function of
$\omega \tau_\eta$ and of $\omega /N$ 
in Fig.~\ref{fig:spctr_a_par}a and b respectively.
The spectra are again non-dimensionalized by $\langle \epsilon \rangle$.
Unlike in the case of $E^a_\perp$, we observe that deviations
from the HIT run are already prominent even for runs with weak
RaS.
All spectra are characterized by the presence of a peak, with
the peak becoming more sharp and prominent as $Fr$ decreases.
An inspection of spectra in Fig.~\ref{fig:spctr_a_par}b clearly shows
that all these peaks correspond to $\omega/N \approx 1$, i.e., the respective 
Brunt-V\"ais\"al\"a frequencies, conforming with the dominant
role of stratification in the vertical direction. 
Note the peak for the stratification-only run is even more
pronounced. 
For runs with strong RaS, i.e., $Fr\le0.069$,
a very minor enhancement of the spectra is visible in the band of frequencies
$\omega/N \approx 0.2$ (or $\omega/f \approx 1$).

Another point to note is that even though
the peaks in $E^a_\|$ become sharper with decreasing $Fr$,
their amplitudes do not vary much, in
contrast to $E^a_\perp$, where the peaks vary
by more than an order of magnitude.
This can be explained by the variation of
acceleration variance as seen in Fig.~\ref{fig:a_0}.
While the $a_0^\perp$ sharply shoots up for
runs with $Fr\le0.069$, the $a_0^\|$ 
only shows a minor variation in comparison.

The results of this Subsection complement the conclusions of
Subsections~\ref{subsec:acc_var} and \ref{subsec:PDF_acc}, as well
as those of Section~\ref{sec:euler}, which were
pointing to two qualitatively different regimes, dominated by waves for 
$R_{IB} < 1$, and by eddies $R_{IB} > 1$. Whereas in the latter regime, 
the variances and the PDFs of accelerations were showing only moderate
deviations compared to HIT flow, Fig.~\ref{fig:spctr_a_perp}b 
and \ref{fig:spctr_a_par}b reveals the role of $N$ and $f$ 
in the horizontal and vertical motion closer to the transition, 
when $R_{IB} \gtrsim 1$
($R_{IB} = 2.69$ for run 3). The transition
to a wave-dominated regime for $R_{IB} < 1$ corresponds to 
the formation of peaks, which are much more intense 
(see Fig.~\ref{fig:spctr_a_perp}b) or much sharper 
(see Fig.~\ref{fig:spctr_a_par}b), leading to a qualitatively 
very different dynamics.

\section{Lagrangian velocity and dispersion statistics}
\label{sec:velocity}

The spread or dispersion of a material under the action of turbulence
is of obvious importance and can be directly studied from
the investigation of the Lagrangian velocity along trajectories.
In this regard, we briefly summarize the classical theory below \cite{MY.I}.

Single-particle dispersion is best understood by considering
the mean-square displacement of a particle from its initial 
position. From Eq.~\eqref{eq:dxdt}, we can write for
each direction:
\begin{align}
\langle Y_i^2 (t) \rangle = \langle u_i(0)^2\rangle 
\int_0^t \int_0^t \rho^{u_i} (  t' ,  t'' ) dt' dt''
\label{eq:disp_1}
\end{align}
where ${Y_i}(t) = {x_i}(t) - {x_i}(0)$ is the displacement 
of the particle from its position ${x_i}(0)$ at $t=0$ to 
its position ${x_i}(t)$ at time $t$,
and $\rho^{u_i}(t', t'')$ is the velocity autocorrelation.
For statistical stationarity, the autocorrelation function 
at $t'$ and $t''$ only depends on $ \tau = | t' - t''|$, allowing Eq.~\eqref{eq:disp_1} 
to be rewritten as
\begin{align}
\langle {Y_i}^2 (t) \rangle = 2 \langle {u_i}^2\rangle 
\int_0^t (t-\tau) \rho^{u_i} (\tau) d\tau
\label{eq:disp}
\end{align}
Once again, we treat separately the 
horizontal and vertical components, as explained before. 
The expression for the dispersion, 
Eq.~\eqref{eq:disp}, simplifies in the 
limit of short and long times, respectively to
\begin{align}
\langle Y_{\perp, \|}^2 (t) \rangle &= \langle u_{\perp, \|}^2\rangle t^2  &\text{for} \ \ 
t \ll \tau_K  \ , \label{eq:ballistic} \\
\langle Y_{\perp, \|}^2 (t) \rangle &= D_{\perp, \|} t    &\text{for}  \ \  t\gg T_L^{\perp,\|}  \ ,
\label{eq:Taylor}
\end{align}
where $D_{\perp, \|}=2\langle u_{\perp, \|}^2\rangle T_L^{\perp,\|}$ are the 
diffusion coefficients
and 
\begin{align}
T_L^{\perp, \|} = \int_0^\infty \rho^{u}_{\perp, \|} (\tau) d\tau
\label{eq:tint}
\end{align} 
are
the Lagrangian integral times.
In the following subsections, we discuss separately 
the behavior of the autocorrelation
functions, integral time scales  and then the mean-square displacements.

\subsection{Velocity autocorrelation and integral time scales}

In this subsection, we investigate the behavior of the velocity 
autocorrelations 
and the corresponding integral time scales derived from them.
%(once again considering separately the perpendicular and parallel components).
%In relation to velocity frequency spectrum, an important point 
Similar to the acceleration spectra, the velocity  frequency spectra
can be obtained through a Fourier transform of the velocity autocorrelations.
However, the velocity spectra can also be simply
derived from the acceleration spectra using the simple relation
$E^u(\omega) = E^a(\omega) /\omega^2$ \cite{Yeung89}.
%We recall that the velocity frequency spectrum can
%be simply derived from the acceleration spectrum
%using the relation $E^u(\omega) = E^a(\omega) /\omega^2$ \cite{Yeung89}.
Consequently, we do not discuss the results for $E^u$  in this section
(since they provide similar information as $E^a$). Instead, we have briefly
summarized them in the Supplementary Material \cite{supp}.

\paragraph*{Perpendicular component:}

\begin{figure}[h]
\begin{center}
\includegraphics[width=7.9cm]{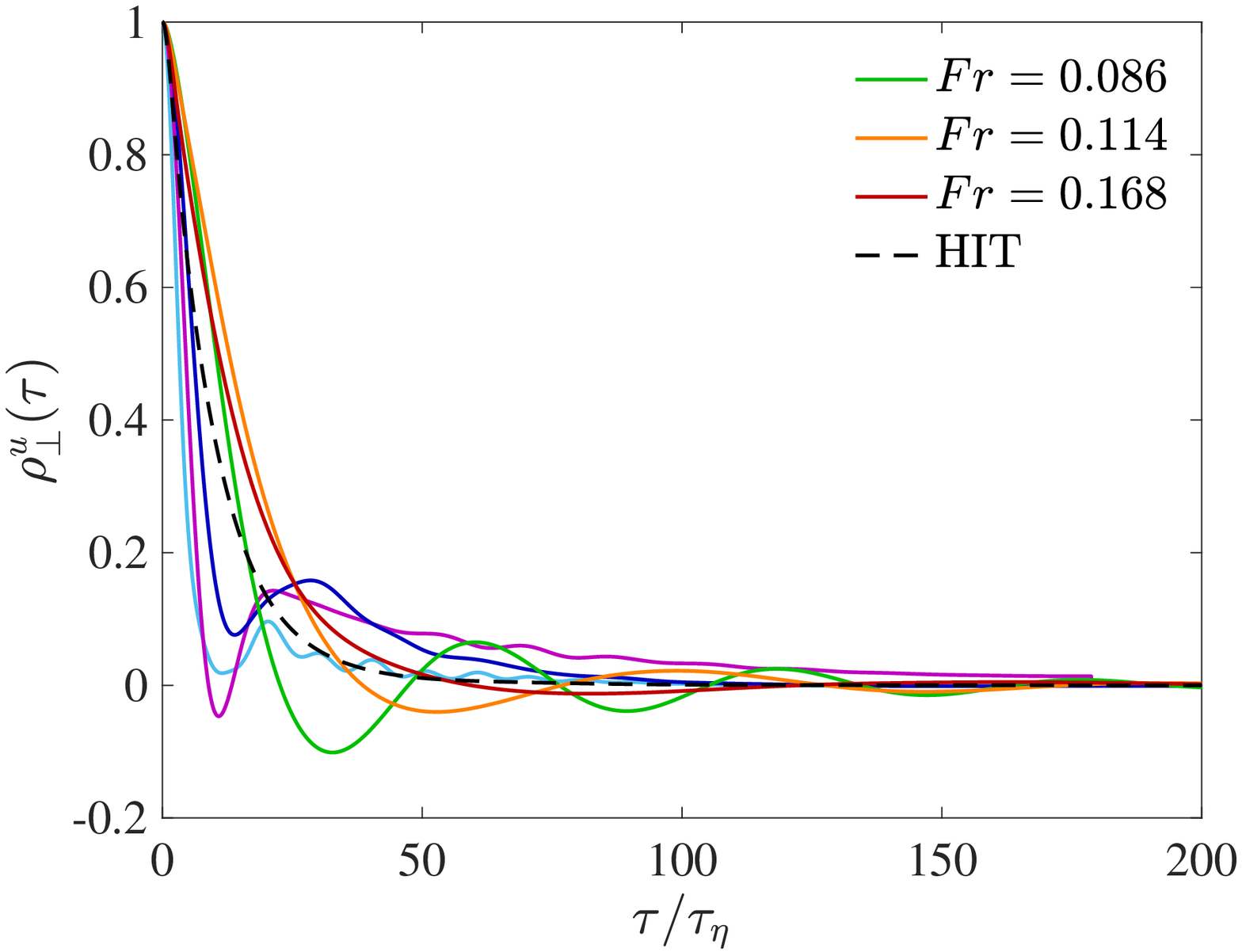}
\includegraphics[width=7.9cm]{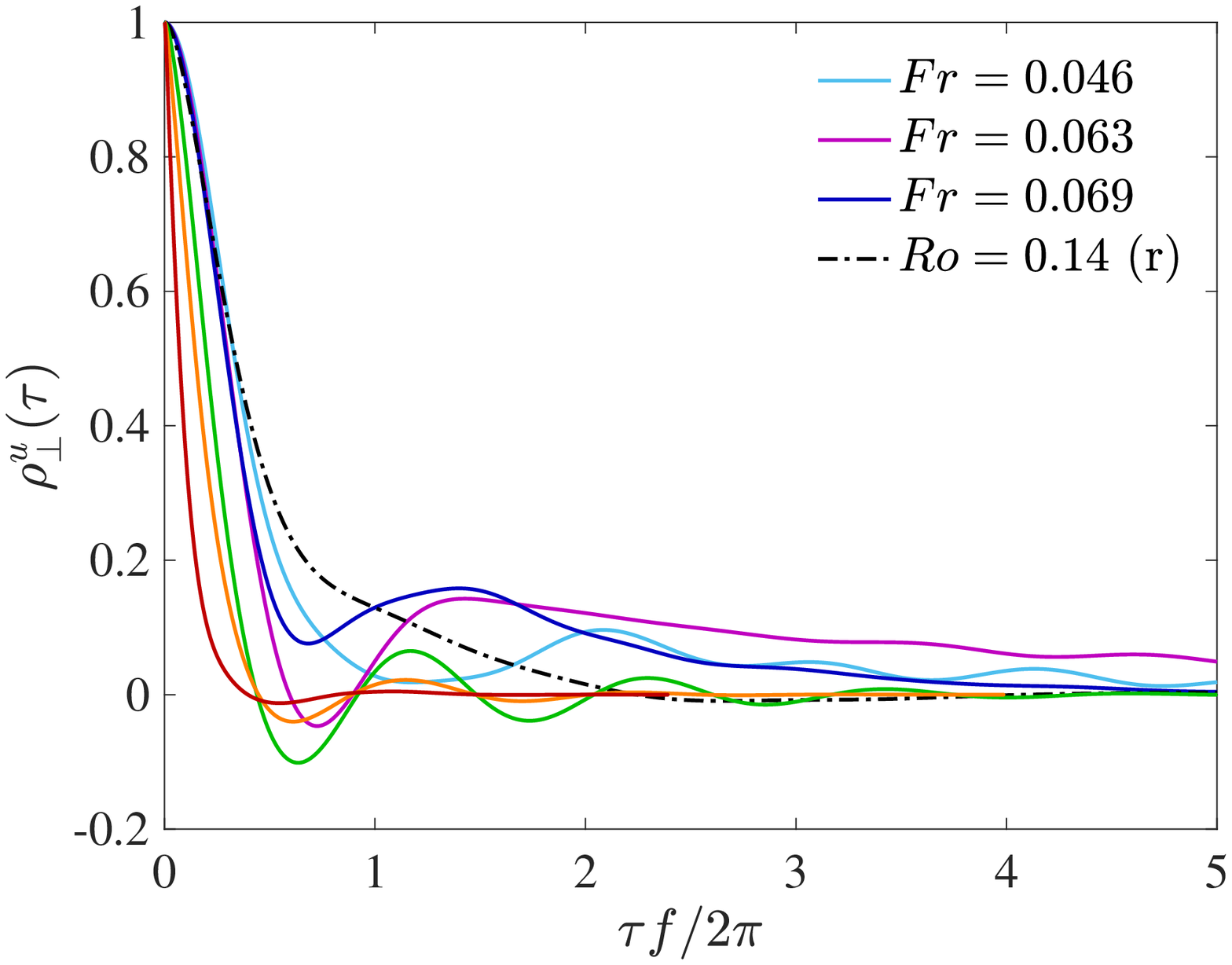}  
\caption{
Lagrangian autocorrelation $\rho^u_\perp (\tau)$ 
of the perpendicular (horizontal) component of
the velocity. Similar normalizations as 
Fig.~\ref{fig:aut_a_perp} are used.
The legend is split over two panels, but applies to each panel individually. 
}
\label{fig:lv_perp}
\end{center}
\begin{picture}(0,0)(0,0)
\thicklines
\put(-175,215){$(a)$}
\put(50,215){$(b)$}
%\put(-165,405){$(a)$}
%\put(60,405){$(b)$}
%\put(-35,225){$(c)$}
%\put(185,225){$(d)$}
\end{picture}
\end{figure}

Fig.~\ref{fig:lv_perp}a-b shows the autocorrelation functions 
$\rho^u_{\perp}(\tau)$;
%along with the corresponding frequency spectra $E^u_{\perp}(\omega)$ in 
%Fig.~\ref{fig:lv_perp}c-d.
similar to results for the perpendicular component of acceleration, 
the time lag $\tau$ is normalized
first by $\tau_\eta$ and then by $f$.
The correlation function $\rho^u_{\perp}$, plotted
as a function of $\tau/\tau_\eta$, see Fig.~\ref{fig:lv_perp}a,
shows only weak deviations from
the HIT case (shown in black dashed line) 
when RaS are moderate (runs 1-3). 
We recall that in the HIT case,
the long time behavior of the velocity autocorrelation can be well
represented by an exponential functional form, 
$ \rho^u \sim \exp^{-\tau/T_L}$ \cite{Yeung02}. 
We notice that the runs with RaS are fundamentally
different from the HIT run with appearance of oscillations resulting 
in negative values of the correlation function. 
These deviations become stronger with imposed RaS.
%In particular, the runs with RaS are fundamentally
%different with appearance of oscillations resulting in negative values 
%of the correlation function. 

To better understand these oscillations, Fig.~\ref{fig:lv_perp}b  
shows the autocorrelation as a function of $\tau f/2\pi$.
Although an oscillatory behavior can be seen in the runs 
with the highest values of RaS rates, the frequency
is close to, but differs from $f$.
On the other hand, 
$\rho^u_\perp$ in the flow
with rotation only (shown in black dash-dotted line 
in Fig.~\ref{fig:lv_perp}b), appears to be monotonically decreasing
(no oscillations).

%Further insight can be obtained from the frequency spectra,
%shown in Fig.~\ref{fig:lv_perp}c-d. 
%Fig.~\ref{fig:lv_perp}d
%shows the spectra as a function of $\omega/f$, and reveals, in the
%presence of RaS,
%the presence of a weak maximum, in a band of frequencies 
%around $\omega/f\approx 1$.
%This differs from the acceleration auto-correlation function, 
%where the rotation frequency was 
%more prominent with increasing strength of RaS. 

\paragraph*{Parallel component:}
%\label{subsubsec:v_par}

\begin{figure}[h]
\begin{center}
\includegraphics[width=7.9cm]{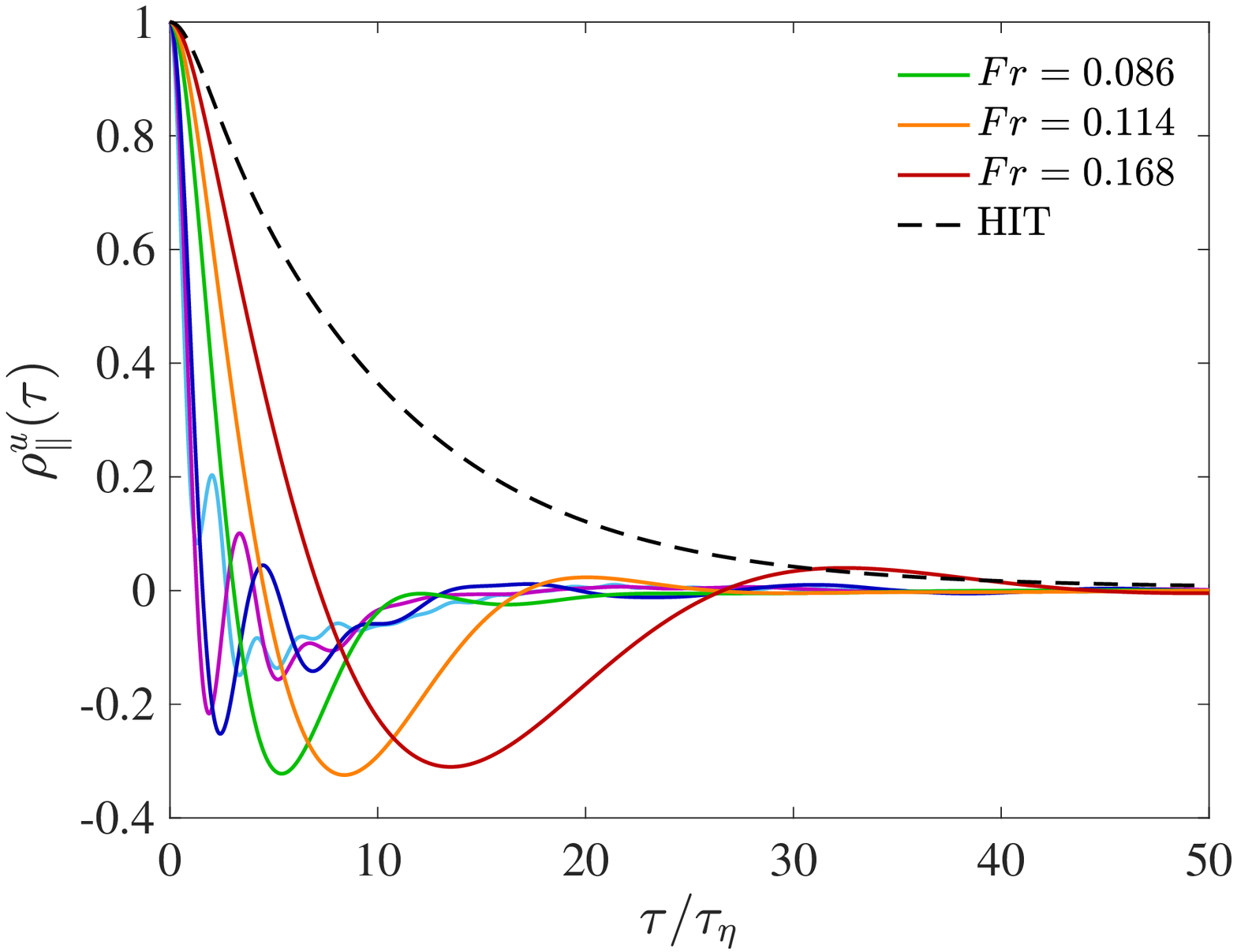}
\includegraphics[width=7.9cm]{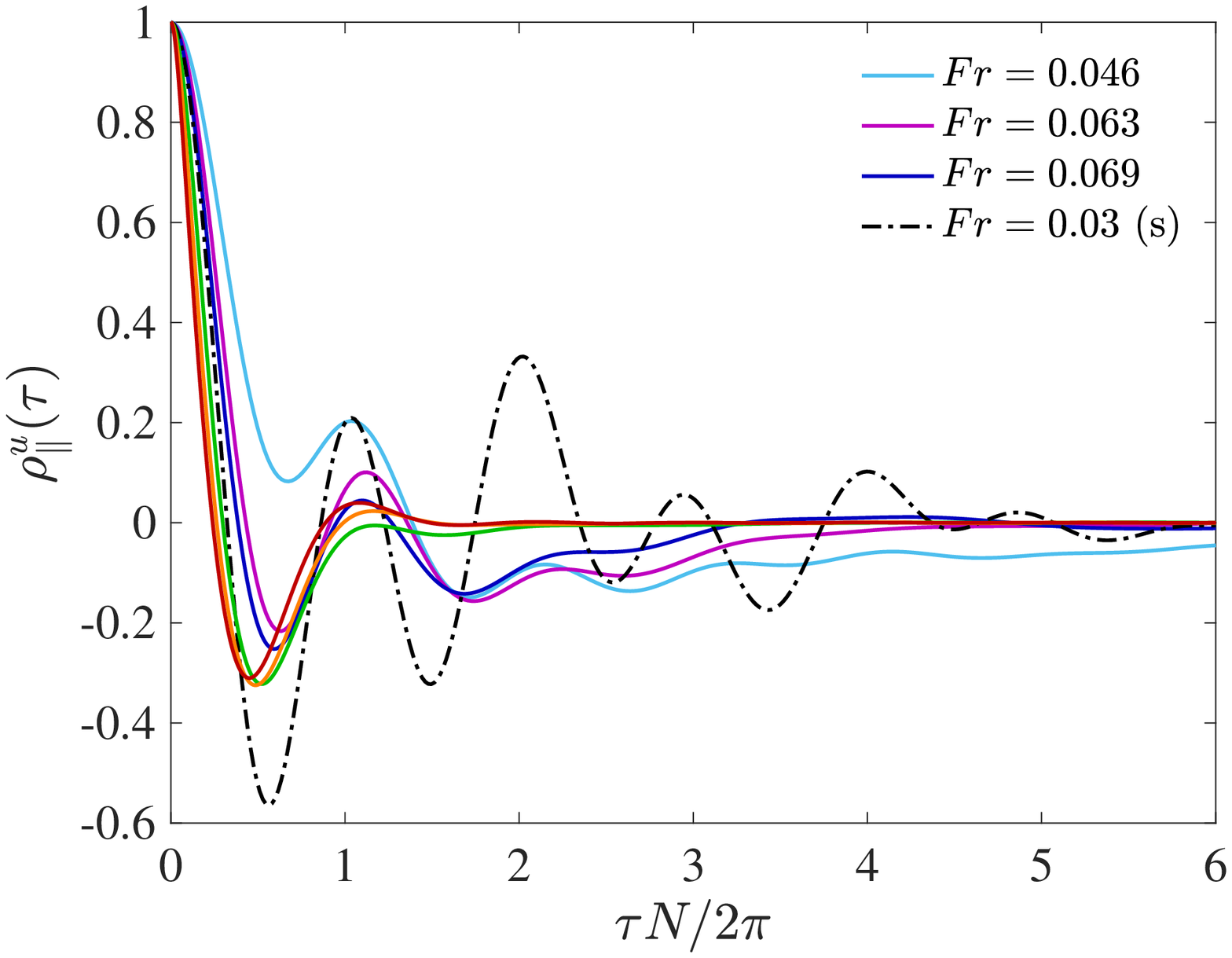} 
\caption{
Lagrangian autocorrelation $\rho^u_\| (\tau)$ 
of the parallel (vertical) component of 
the velocity. Similar normalizations as 
Fig.~\ref{fig:aut_a_par} are used.
The legend is split over two panels, but applies to each panel individually. 
}
\label{fig:lv_par}
\end{center}
\begin{picture}(0,0)(0,0)
\thicklines
\put(-175,215){$(a)$}
\put(50,215){$(b)$}
%\put(-165,405){$(a)$}
%\put(60,405){$(b)$}
%\put(-35,225){$(c)$}
%\put(185,225){$(d)$}
\end{picture}
\end{figure}

The autocorrelations and spectra corresponding to the parallel
component of velocity are shown in Fig.~\ref{fig:lv_par}.
We first consider the autocorrelation $\rho^u_\|$  as a function of
$\tau/\tau_\eta$ in Fig.~\ref{fig:lv_par}a. 
In all the cases with RaS, 
$\rho^u_\|$ strongly differs from the HIT case 
(indicated by a black dashed line).
Two patterns are distinctly visible. First, all the autocorrelation
functions
decay rapidly. In terms of the dimensionless time $t/\tau_\eta$, the
decay rate becomes increasingly larger when $Fr$ decreases
(especially in comparison to perpendicular component
shown in Fig.~\ref{fig:lv_perp}a).
Second, the autocorrelations all overshoot to become strongly 
negative and show distinct oscillations.
The primary period of these oscillations corresponds to the stratification 
frequency, as demonstrated by
Fig.~\ref{fig:lv_par}b, which shows the autocorrelations
as a function of $\tau N/2\pi$. In addition, Fig.~\ref{fig:lv_par}b
also shows that all cases superpose reasonably well for the initial
decay of the autocorrelation,
These results clearly demonstrate that stratification plays a dominant
role in the vertical motion, and in comparison
turbulence plays a weaker role. 
The fast dynamics in the vertical direction 
is dominated by the strong vertical gradients 
arising from the constraint that the vertical Froude number be of 
order unity \cite{billant_01}, 
and leading to strong spatial and temporal intermittency, 
dissipation and mixing \cite{pouquet_19p}.
%turbulence plays a minor role. 
This is true, even for the runs at weak RaS
($Fr \gtrsim 0.08$). A similar behavior was obtained for
the vertical acceleration correlation function, see
Section~\ref{subsec:Corr_acc}. 
We notice that in the case of a purely
stratified run, without any rotation, the autocorrelation function 
$\rho^u_{\|}$ clearly oscillates, with a modulation at twice
the Brunt-V\"ais\"al\"a period.

\paragraph*{Integral time scales:}
\label{subsec:tl}

\begin{table}[h]
\centering
    \begin{tabular}{l|c | cccccc |cc}
  \hline \hline
    case      &0    & 1     & 2      & 3     & 4   & 5   & 6 & 7 & 8    \\ 
\hline
    $Fr$   & $\infty$ & 0.168  & 0.114  & 0.086 & 0.069 & 0.063 & 0.045 & 0.030    & $\infty$  \\
    $Ro$   & $\infty$ & 0.840  & 0.570  & 0.430 & 0.345 & 0.315 & 0.225 & $\infty$ & 0.140     \\
\hline
    $T_L^{\perp}/\tau_\eta$  & 11 & 14    & 14    & 10   & 12   & 4.1   & 5.8   & $\infty$    & 8.2       \\
    $T_L^{\|}/\tau_\eta$     & 10 & 0.59    & 0.23    & 0.11   & $6.4 \, 10^{-2}$   & $2.7 \, 10^{-2}$   & $5.5 \, 10^{-2}$  & $3.6 \, 10^{-2}$       & 8.4       \\
\hline
    $T_L^{\|} N/( 2 \pi)$    & 0     & $2.0 \, 10^{-2}$    & $1.4 \, 10^{-2}$    & $1.1 \, 10^{-2}$   & $1.6 \, 10^{-2}$   & $9.0 \, 10^{-3}$   & $2.8 \, 10^{-2}$   & $1.2 \, 10^{-2}$       & 0       \\
\hline
    \end{tabular}
\caption{
Integral time scales
of the velocity, $T_L^{\perp}$ and $T_L^{\|}$ respectively
based on the perpendicular (horizontal) component and parallel (vertical) component 
(see Eq.~\eqref{eq:tint} for definition).
The integral time scales are also compared with the Kolmogorov time $ \tau_\eta$, 
and of the Brunt-V\"ais\"al\"a period, $2 \pi/N$, for the vertical motion, and 
of the rotation period, $ 2\pi/f$, for the horizontal motion.
(A more detailed table is provided in the Supplementary Material \cite{supp}).
}
\label{tab:tl}
\end{table}

As highlighted by Eqs. \eqref{eq:Taylor} and \eqref{eq:tint},
the Lagrangian integral time scale  $T_L$,
characterizes the particle dispersion at large times and
is simply obtained by integrating the autocorrelation function:
$T_L = \int_0^\infty \rho^u d\tau$.
We begin by noticing that many of the correlation functions shown in 
Figs.~\ref{fig:lv_perp} and \ref{fig:lv_par} exhibit an oscillatory behavior.
This makes the calculation of their integrals generally prone to statistical
errors. As such, the values of $T_L^{\perp,\|}$, listed in 
Table~\ref{tab:tl}, are subject to 
large relative errors, the more so as the values of $T_L^{\perp,\|}$ 
are small.

Over the range of $Re$ considered here,
the variation of $T_L^{\perp}$ 
is found to be rather weak at moderate RaS ($Fr \gtrsim 0.08$).
In particular,
the ratio $T_L^{\perp}/\tau_\eta$, is of the order of $10$
for the HIT run, and remains approximately constant
for cases 1-4 and also for the run 8, which corresponds to a flow with
rotation only.
For cases 5-6 with strong RaS, the ratio sharply drops by a factor $\sim 2$,
which can be attributed to a sharp decrease in the
mean dissipation $\langle \epsilon \rangle$ (hence causing a
sharp increase in $\tau_\eta$).
The above observations are consistent with earlier 
results which also highlight a sharp transition
to wave-dominated regime as $R_{IB}$ significantly decreases.
We note that for the run with pure stratification (case 7), 
the {velocity} autocorrelation in the horizontal direction 
{converges extremely slowly}
to zero at large time lags, pointing to an extremely large value
of $T_L^{\perp}$. A similar behavior has been 
noticed in~\cite{Sujovolsky:18}.
In this regard, the presence of rotation plays a crucial 
role. In particular, the 
dispersion relation and the structure of the eigenmodes, underlying the
wave motion
is strongly modified by rotation.

On the other hand, 
the integral time $T_L^{\|}$ is  
substantially smaller than $T_L^{\perp}$
for all runs with stratification (cases 1-7).
The ratio $T_L^{\|}/\tau_\eta$ is still around $10$
for the HIT and purely rotating flows (cases 0 and 8 respectively),
but is greatly reduced for all runs with stratification.
This can be evidently attributed to the strongly oscillating nature
of the velocity autocorrelation functions for cases 1-7
(see Fig.~\ref{fig:lv_par}b), which results in significant 
cancellation when calculating $T_L^\|$.
The ratio $T_L^{\|}/\tau_\eta$ drops by more than 
an order of magnitude in going from HIT to case 1, which corresponds
to the weakest RaS. Thereafter, with 
increasing strength of RaS, the 
ratio further decreases.  
Interestingly, the ratio $T_L^{\|}N/2\pi$
shows a much reduced variation, and remains in the range 0.01-0.02
for all cases with stratification. This suggests
that despite the strong oscillations in the autocorrelation
the time scale, $T_L^\|$, is not strictly zero, but 
instead scales inversely with $N$, consistent with earlier
predictions for purely stratified flows \cite{lb08}.

\subsection{Mean-square displacement}
\label{subsec:displ}

\begin{figure}[h]
\begin{center}
\includegraphics[width=7.9cm]{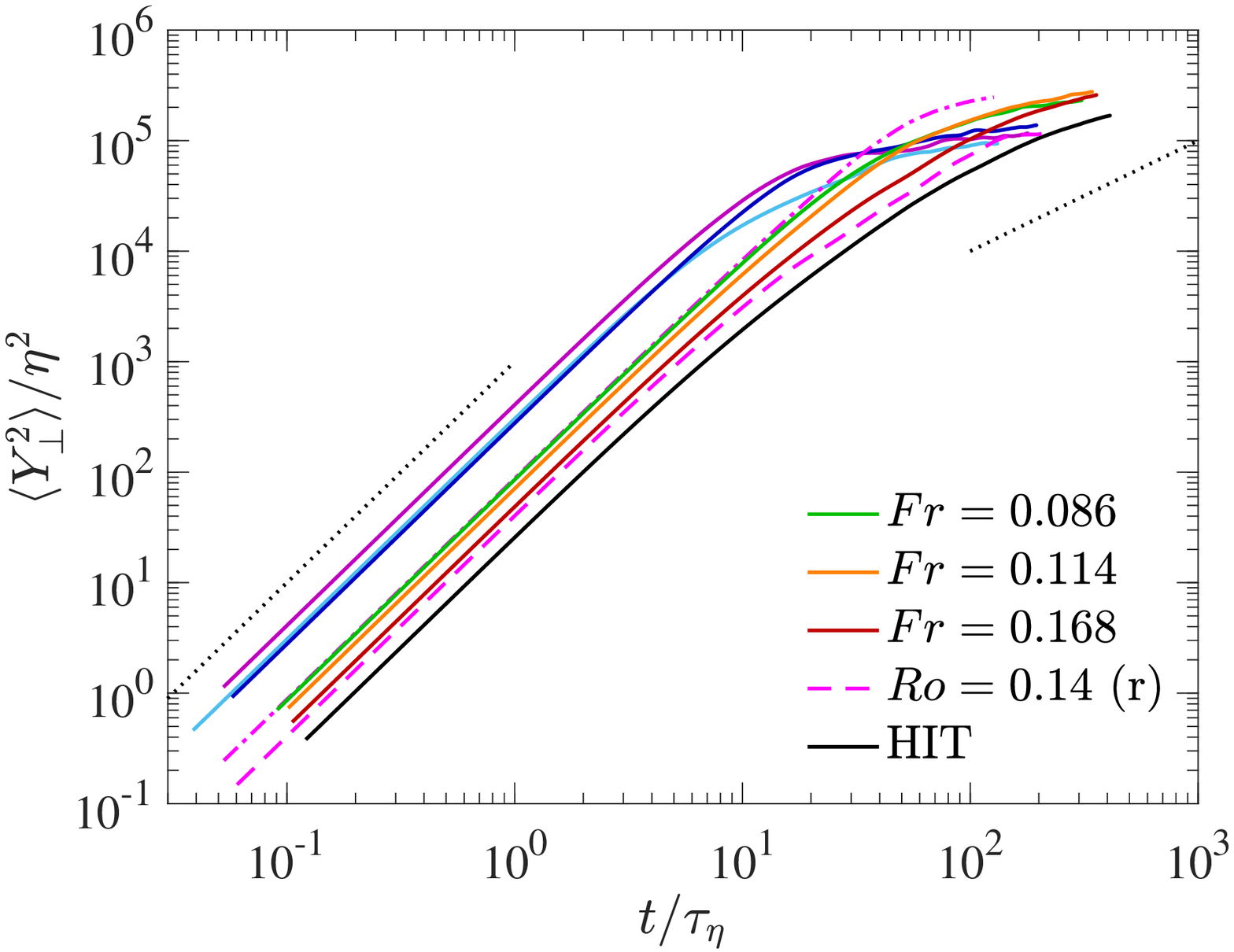}
\includegraphics[width=7.9cm]{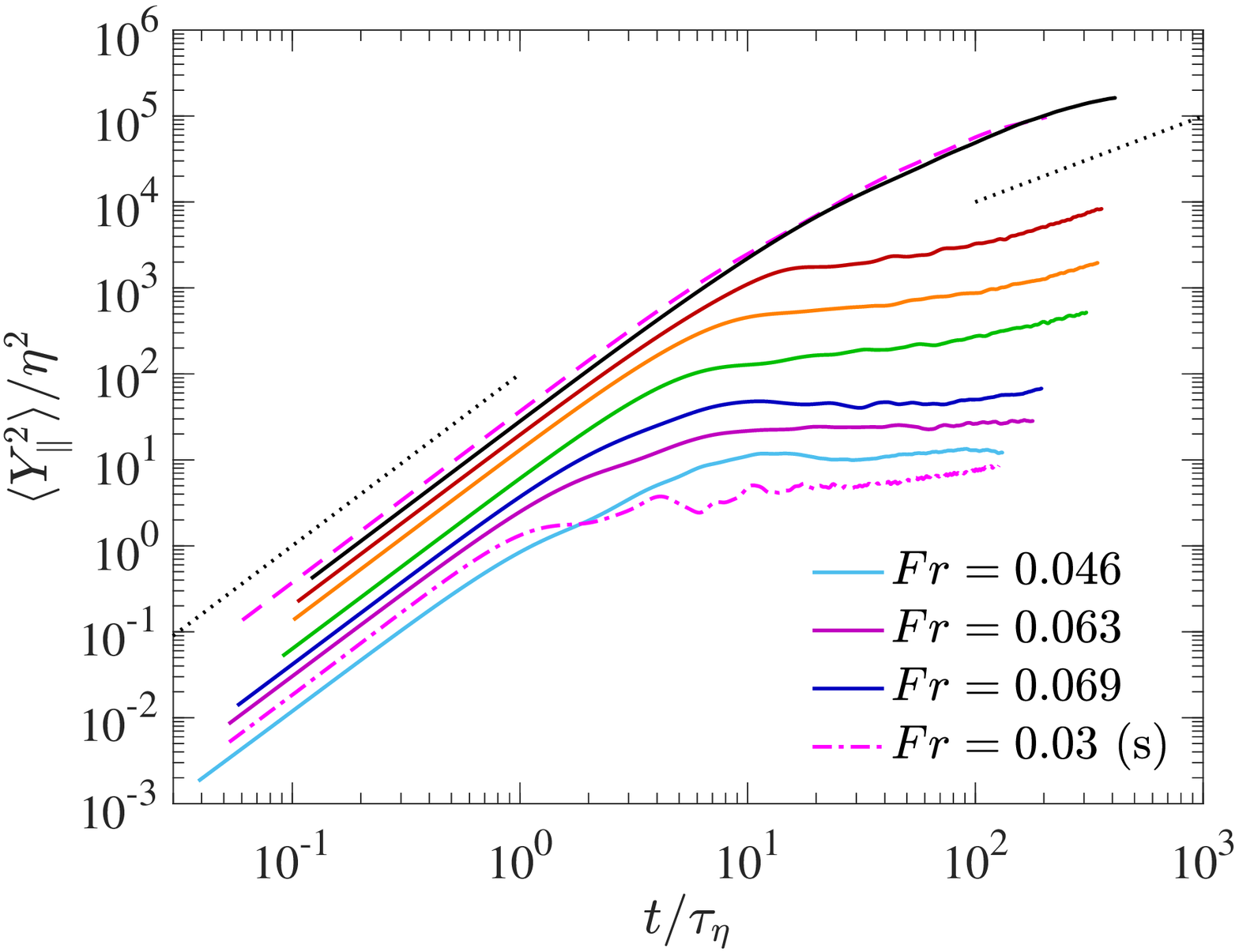}
\caption{Mean square displacement as a function of time 
in the (a) perpendicular (horizontal) direction
and (b) parallel (vertical) direction. 
All quantities normalized by Kolmogorov scales.
The dotted lines
represent slopes 2 and 1 at short and long times respectively.
The legend is split over two panels, but applies to each panel individually. 
}
\label{fig:disper}
\end{center}
\begin{picture}(0,0)(0,0)
\thicklines
\put(-165,215){$(a)$}
\put(60,215){$(b)$}
\end{picture}
\end{figure}

Fig.~\ref{fig:disper} shows the mean-square displacement
for horizontal and vertical directions as a function of time
(both axes normalized appropriately by
Kolmogorov scales).
At small times, a clear $t^2$ scaling is visible 
for all runs in both components, as expected from 
Eq.\eqref{eq:ballistic}. 
The displacement in the horizontal 
direction is enhanced by RaS,
whereas inhibited in the vertical direction.
This can be readily explained by considering the
variance of the components of the
velocity $\langle v_\perp^2 \rangle$ and $\langle v_\|^2 \rangle$, as shown 
in Table~\ref{table:stat_v} (see also Eq.~\eqref{eq:ballistic}).

At long times, 
transition to a diffusive regime, given by 
Eq.~\eqref{eq:Taylor}, is expectedly seen in the HIT run ~\cite{Yeung89},
for both horizontal and vertical components.
A similar behavior is also visible for runs with
weak RaS ($Fr \gtrsim 0.08$), when considering the 
horizontal component. 
In comparison, the growth 
of the vertical component at long times for these runs is significantly inhibited,
nevertheless still showing a very slow approach to a diffusive regime.
Our results are generally consistent with those obtained earlier for a
comparable parameter range ($N/f = 10$, $Fr \approx 0.16$)~\cite{cambon:04}.
Note the rotation-only run, behaves like the HIT run,
consistent with the small degree of anisotropy in
both velocity and acceleration statistics as noted earlier.

On the other hand, for runs with strong RaS ($Fr \lesssim 0.07$), 
the growth of both horizontal and vertical components appears to be
slower than a linear behavior.  
This is particularly clear for the vertical displacements
$\langle Y_{\|}^2 \rangle$, which appear to increase extremely slowly
with time, if at all.  
In fact, the values of $\langle Y_\|^2 \rangle/\eta^2$ remain 
small, of the order of $\sim 10$, over a very long time.
The absence of a clear diffusive regime for $\langle Y_\|^2 \rangle$ is
arguably not very surprising, given the small values of $T^L_\|$ and 
$\langle v_\|^2 \rangle$, hence of the diffusion coefficient 
$D = \langle v_\|^2 \rangle T_L^\|$. 
Assuming that  $T^L_\|$
is inversely proportional to $N$ (multiplied with a small prefactor), 
the expected
diffusive behavior will likely be visible only after an extremely 
long time. Consistent with our own results, 
the near constancy of $\langle Y_\|^2 \rangle$ was
also reported in purely stratified flows 
\cite{Artrijk:08,Brethouwer:09,Sujovolsky:18}.

Contrary to the very low values of $T_L^{\|}$ for runs with strong RaS
(see Table~\ref{tab:tl}), the values of $T_L^{\perp}$ do not 
appear to be particularly small. This makes the lack of a clear diffusive regime 
for runs 4-6 for the horizontal displacement much more surprising.
It has been noticed in the context of stratified flows,
that the horizontal displacement could grow as $t^2$ \cite{Artrijk:08}. 
Here, our results show that, 
in a regime dominated by waves (runs 4-6), the growth of 
$\langle Y_\perp ^2 \rangle$ is slower than $t^1$ over a long time.
This suggests a strong interaction between the horizontal and vertical 
directions owing to the presence of rotation 
(and inertia-gravity waves), which was also visible
in autocorrelations of acceleration and velocity. 
However, given the large values of $T_L^{\perp}$,
it can be expected that the diffusive regime is ultimately reached
at sufficiently longer times.

Finally, we note that the strong vertical drafts
responsible for high kurtosis of $u_\|$ (see Section~\ref{sec:euler}), 
and increased mixing efficiency \cite{Feraco:18}, 
do not appear
to cause an appreciable growth in $\langle Y_\| ^2 \rangle$
for the runs considered here.
This is not inconsistent given the 
kurtosis is a normalized fourth-order moment,
whereas $\langle Y_\| ^2 \rangle$  
goes as the variance of vertical velocity
and the integral scale $T_L^{\|}$,
both of which are strongly suppressed.
It is likely that at higher Reynolds numbers than considered
here, the effect of the vertical drafts is felt more directly
on the vertical displacement of fluid particles.

\section{Discussion and Conclusions}
\label{sec:concl}

Motivated by the well established effect of both 
rotation and stratification (RaS) 
in numerous geophysical flows,
and by the observation
of Lagrangian tracers, such as buoys 
in the ocean \cite{poje_17},
% or balloons over Antarctica \cite{walterscheid_16,Maruca_17},
we have investigated the properties of particle trajectories in such flows.
We have utilized direct numerical simulations of the 
Boussinesq equations in a periodic domain,
with both stratification and the axis of rotation aligned in the vertical direction, 
in a statistically stationary regime maintained by an external forcing.
We have focused on cases where the corresponding rotation and stratification
frequencies,  $f$ and $N$ respectively, are held in a fixed ratio $N/f=5$,
as relevant in oceanographic situations \cite{Nikurashin:12}.
The usual computer limitations have restricted the Reynolds numbers
to approximately around $4000$ in our runs,
which are smaller than those observed in nature,
but still large enough to allow turbulence to adequately develop.
The Froude numbers investigated are in the range 
$0.03 \lesssim Fr \lesssim 0.2$,
with corresponding Rossby number $Ro=5Fr$, compatible with
geophysical fluid flows.
Our results illustrate the complex physical effects involved 
in a turbulent flow under the combined action of RaS. Whereas the
corresponding physical effects have been recently investigated in an Eulerian 
context in many studies, we have documented 
the elementary properties of particle trajectories,
in particular concerning acceleration, velocity and displacement
statistics.
Due to the imposed large-scale anisotropy, 
we differentiate between motion in the horizontal plane and in the vertical direction.
We also appropriately compare with rotation-only and stratification-only runs and 
also with the well studied case of isotropic turbulence.

While the parameters $Fr$ and $Ro$  
vary smoothly with 
the strength of imposed RaS, we observe a sharp change in the nature of
the flow when $Fr \lesssim 0.07$ for the runs considered,
corresponding to a regime where waves appear to dominate over the
non-linearities, as also observed in \cite{Feraco:18} for stratification-only runs.
An inspection of the Eulerian energy spectra (and the temperature spectra),
shows a clear transition from a Kolmogorov $k^{-5/3}$ like behavior
to a Bolgiano $k^{-11/5}$ like behavior at intermediate wavenumbers.
The transition also corresponds to the peak of intermittent disruption 
of shear layers, 
with a maximum of the kurtosis of the vertical velocity. In this 
context, our work extends the stratification-only results of \cite{Feraco:18}. 
Interestingly, we find that this transition to wave-dominated regime
can be given by a simple condition of 
$R_{IB}<1$ based on the buoyancy Reynolds number $R_{IB}$.

The effect of this transition is also evident in Lagrangian statistics.
For acceleration statistics, the effect of imposed RaS appears very weak before the
transition, suggesting the imposed anisotropy does not affect the small-scale isotropy
significantly. The probability density functions (PDFs) of acceleration
in both horizontal and vertical direction display wide tails very similar 
to isotropic turbulence, 
and the variance of acceleration can be approximately scaled with 
Kolmogorov variables. 
The Lagrangian autocorrelations and frequency spectra also appear 
to behave in a similar fashion, although 
a weak effect of RaS is still visible at low frequencies
in the spectra (when $R_{IB} \gtrsim 1$).
%the frequencies $N$ and $f$ become visible in the spectra of $a_\perp$ and $a_\|$, 
However, the sharp transition at $R_{IB} \lesssim 1$ leads to 
a very strong anisotropy between the horizontal and vertical directions,
with the particle motion strongly dominated by RaS instead of turbulent eddies.
The PDFs of acceleration show suppressed tails,
and acceleration variance shows no clear scaling.
Moreover, the autocorrelations and frequency spectra clearly reflect
the dominant roles of frequencies $N$ and $f$.

On the other hand, Lagrangian velocity statistics are always
affected by the imposed RaS. Anisotropy is already evident
for weak RaS; however the degree of anisotropy becomes very strong
with the said transition at $R_{IB} \lesssim 1$. 
This is most readily seen in the integral time scale based on the 
autocorrelations from vertical velocity
and hence also reflected in long time behavior of 
mean-square displacement of particles in the vertical direction.
Similar effects had been observed in the 
case of purely stratified flows~\cite{Sujovolsky:18}. In such flows, it 
is known that stratification leads to the formation of horizontal layers, that
inhibits the 
transport in the vertical direction. 
In the presence of both RaS, slanted layers appear \cite{marino13b} 
which somewhat inhibit transport in both the vertical and the horizontal directions. 

It is interesting to consider that the transitional behavior observed 
in this work is possibly also
linked to the enhanced vertical velocities that develop in 
strongly stably stratified turbulent flows, such as in the nocturnal 
planetary boundary layer \cite{lenschow_12}, and warrant further examination.
The resulting dispersion of Lagrangian particles must feel these hot spots of 
strong vertical velocity and strong local dissipation linked to localized shear instabilities, 
as also observed in the ocean.
Several studies point to a local Richardson number, measuring 
stratification with respect to the internal shear associated with 
vertically sheared horizontal winds, to be mostly in the vicinity of the
value for linear instability \cite{smyth_19,Rosenberg:15,pouquet_19p}.
This leads to flows which are altogether close to criticality, and strongly
anisotropic, intermittent and dissipative \cite{pouquet_19p}. 
We expect that the observed similarity between the statistical
properties of acceleration when the flow is dominated by eddies
(for $R_{IB} > 1$), and those in HIT, may be very useful in modeling 
the dispersion of particles in turbulent flows in the presence of RaS.

Whereas %the Reynolds number 
$R_{IB}$ seems to
provide a clear criterion to distinguish between the different regimes 
observed here, alternative ways~\cite{pouquet_18} have been proposed to 
describe the transition 
between wave-dominated flows, and flows where waves and eddies interact. 
Namely, it was found that the transition
occurs around $R_{\cal B} \sim 10$, where $R_{\cal B}=Re Fr^2$. In this sense, the flows studied here
are all in the regime of interacting waves and eddies, although run 6 is very
close to the transition towards a wave-dominated flow.

%we stress that the
%picture we are drawing 
%is based on runs at a
%relative moderate
%Reynolds numbers.
%% $ 3,000 \lesssim Re \lesssim 6,000$.
%A natural question arises about 
%whether these conclusions would apply at even larger Reynolds numbers.
%Further investigations, varying not only the Reynolds number, but also the 
%ratio $N/f$, should bring more insight on the structure of these flows.

We conclude by mentioning that many interesting issues remain to be 
addressed for particle motion in such flows.
For example, the problem of separation of pairs, or of clusters of 
particles, both forward and backward in time ~\cite{BSY.2015, Xu2015}
or considering the effects of particle inertia or molecular diffusion.
In addition, 
we stress that the picture we have drawn is based on runs at a
relative moderate Reynolds numbers.
Exploring higher Reynolds numbers, and also higher Prandtl/Schmidt numbers \cite{buaria20}, 
along with different values of
$N/f$ will help in better understanding the parameter space and 
how the results from current work translate
to practical geophysical applications.
% \cite{marino_15p, pouquet_18}.

\section*{Acknowledgments}

The computing resources utilized in this work
were provided by
PSMN at the Ecole Normale Superieure de Lyon.
D. Buaria and A. Pumir acknowledge support from 
the European High-performance Infrastructure in Turbulence
(EuHIT) program and 
the Max Planck Society. 
A. Pumir acknowledges additional support
from the IDEXLYON project (Contract ANR-16-IDEX-0005)
under University of Lyon auspices. 
R. Marino acknowledges support from the program PALSE
(Programme Avenir Lyon Saint-Etienne) and IMPULSION of the 
University of Lyon, in the framework of the program
{\em Investissements d'Avenir} (ANR-11-IDEX-0007),
from the Inge'LySE (Lyon-Saint-Etienne) and from ERCOFTAC.
A. Pouquet gratefully acknowledges support from 
Laboratory for Atmospheric and Space Physics (LASP), 
in particular Bob Ergun, at the
University of Colorado, Boulder.
%We also thank the anonymous for their comments.
%\{I would not go this way !!!}

%\bibliography{references}


\end{document}